\documentclass[12pt]{spieman}  
\usepackage{amsmath,amsfonts,amssymb}
\usepackage{graphicx}
\usepackage{setspace}
\usepackage{tocloft}
\usepackage{lineno}
\usepackage{color,soul}
\usepackage{verbatim}

\title{Design of the tertiary optical system for the LLAMA radio telescope}

\author[a,c,*]{Emiliano Rasztocky}
\author[b,d]{Matías Rolf Hampel}
\author[e]{Rodrigo Reeves}
\author[f]{Jacques R.D. Lepine}
\author[a,g]{Gustavo Esteban Romero}
\affil[a]{Instituto Argentino de Radioastronomía (CONICET - UNLP - CIC), Berazategui, Argentina}
\affil[b]{Instituto de Tecnologías en Detección y Astropartículas (CNEA - CONICET - UNSAM),Buenos Aires,  Argentina}
\affil[c]{Universidad Nacional de San Martín, San Martín Argentina}
\affil[d]{Universidad Tecnológica Nacional - Facultad Regional Buenos Aires, Argentina}
\affil[e]{Universidad de Concepción, CePIA, Departamento de Astronomía, Concepción, Chile}
\affil[f]{Universidade de São Paulo, Instituto de Astronomia, Geofísica e Ciências Atmosféricas, São Paulo, Brazil}
\affil[g]{Universidad Nacional de La Plata, Facultad de Ciencias Astronómicas y Geofísicas, La Plata, Argentina}

\cftpagenumbersoff{figure}
\cftpagenumbersoff{table} 
\begin{document} 
\maketitle

\begin{abstract}
Many modern radio telescopes employ an observational strategy that involves maximizing the use of their available spaces (cabins), outfitting them with various receivers at different frequencies to detect incoming signals from the sky simultaneously or individually. The Large Latin American Millimeter Array (LLAMA), is a joint venture between Argentina and Brazil consisting of the installation and operation of a 12-meter aperture Cassegrain telescope. It features three available cabins for instrumentation and plans to install six single-pixel heterodyne receivers, covering different bandwidths in the 30 to 950 GHz window of the electromagnetic spectrum, in its two lateral Nasmyth cabins at different phases of the project. Therefore, it is crucial not only to design a tertiary optical system that couples the antenna beam to those receivers, but also to do it in a scalable way. The primary goal for the design is to simultaneously maximize the antenna efficiency while minimizing optical aberrations for all receivers, both fundamental aspects for the optimal functioning of cutting-edge astronomical instruments. 
In this paper, we present the entire design process, starting from the quasi-optical approach based on the propagation of a fundamental Gaussian beam mode, continuing with the validation of the design based on physical optics simulations, and ending with a tolerance analysis of the system. 
As a result of this process, a frequency independent tertiary optical system has been achieved for almost all the receivers, which is expected to provide high optical performance for the radio telescope. 
\end{abstract}


{\noindent \footnotesize\textbf{*}Emiliano Rasztocky,  \linkable{eraszto@iar.unlp.edu.ar} }


\section{Introduction}
\label{sect:intro}  

\subsection{Large Latin American Millimeter Array (LLAMA) Radio Telescope}
\label{LLAMA radiotelescope}

\begin{figure}[H]
\begin{center}
\begin{tabular}{c}
\includegraphics[height=6.1cm]{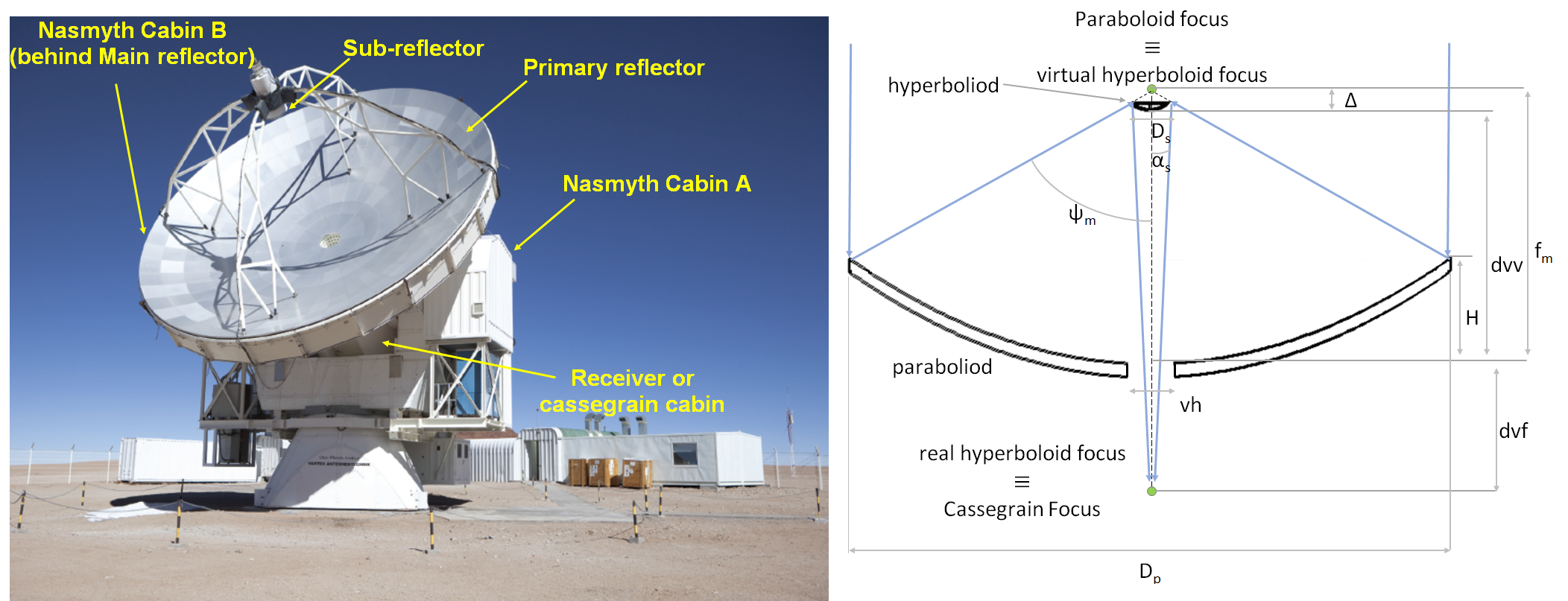}  
\\
(a) \hspace{7.1cm} (b)
\end{tabular}
\end{center}
\caption 
{ \label{fig:APEX_LLAMA_optical_sketch}
(a) APEX antenna at Atacama in Chile. Credit: ESO. (b) Optical configuration of the LLAMA Cassegrain system.} 
\end{figure}

LLAMA \cite{arnal2017llama, llamaweb, romero2020large, lepine2021llama}, is a joint scientific and technological venture between Argentina and Brazil. Its aim is to establish and operate a radio observatory at Alto Chorrillos, Argentina, at an altitude of 4800 m above sea level, capable of making astronomical observations in the millimetre and sub-millimeter wavelength ranges. The main scientific objectives of LLAMA are the study of spectral molecular lines of forming stars, young stellar objects, black holes, the Sun's chromosphere, outflows in starbursts, the chemistry of the intergalactic medium around galaxies and dusty Active Galactic Nuclei (AGN), galaxy formation at high redshifts and even Cosmic Microwave Background (CMB) fluctuations at small angular scales\cite{romero2020large, lepine2021llama, fernandez2023llama}. The radio telescope is a 12-meter aperture Cassegrain-type dual reflector antenna manufactured by Vertex Antennentechnik and shares the same architecture and optical characteristics (Table \ref{table:LLAMA_optical_parameters}) as that of the Atacama Pathfinder Experiment (APEX) radio telescope \cite{gusten2006atacama} (Fig. \ref{fig:APEX_LLAMA_optical_sketch}). The antenna has been designed to provide an absolute pointing accuracy of 2\,arcsec with an offset pointing of 0.6\,arcsec and a reflector surface accuracy of 25\,$\mu$m. These features will allow the observation of the universe in the 30 to 950 GHz window of the electromagnetic spectrum. The radio telescope will be able to operate as a single dish or as part of an interferometer array such as ngEHT\cite{johnson2023key}.

Many modern radio telescopes maximize the use of their available cabins by outfitting them with various receivers operating at different frequencies, allowing both rapid switching between frequencies or even multi-frequency observations. It was decided early in the project that LLAMA, with three cabins - the central Cassegrain cabin (Cab-Cass) and two Nasmyth cabins (Cab-A and Cab-B) - would host six ALMA\cite{almaweb}-like single-pixel heterodyne receivers in the Nasmyth cabins. For the First Light (FL) phase, receivers for bands 5 (163-211\,GHz), 6 (211-275\,GHz), and 9 (602-720\,GHz) will be installed in Cab-B to validate the technical and scientific capabilities of the telescope. Subsequently, a significant upgrade will configure the cabins for a longer-term phase (LT). In this phase, bands 6, 7 (275-373\,GHz), and 9 will be installed in Cab-A, and bands 1 (35-52\,GHz), 2+3 (67-116\,GHz), and 5 in Cab-B. This configuration leaves Cab-Cass available for future upgrades, which might include not only single-pixel detectors but also multi-detector cameras to take advantage of the larger field of view ($\sim$10 arcmin). Implementing this strategy requires using tertiary optical systems to couple the detectors (receivers) with the antenna.

\begin{table}[H]
\caption{Optical parameters of the LLAMA radio telescope \cite{LLAMAmanual} (Fig. \ref{fig:APEX_LLAMA_optical_sketch}).} 
\centering
\begin{tabular}{|lcl|}
\hline
\multicolumn{3}{|l|}{\textbf{Parabolic primary reflector:}} \\                

Diameter & $D_p$= &12\,m \\
Focal length & $f_p$= &4.8\,m \\
Focal ratio & $F_p$= &0.4 \\
Eccentricity & $e_p$= &1 \\
Aperture illumination & $\psi_p$= &64.01\,° \\
Depth of primary reflector & H= &1.875\,m \\
Primary vertex hole diam. & vh= &0.75\,m \\
 & &\\
\multicolumn{3}{|l|}{\textbf{Hyperbolic secondary reflector (sub-reflector):}} \\                                                                                                                              
Diameter & $D_s$= &0.75\,m \\
Focal length & $f_s$= &0.30962\,m \\
Eccentricity & $e_s$= &1.10526 \\
Secondary illumination & $\alpha_s$= &3.58\,° \\
 & &\\
\multicolumn{3}{|l|}{\textbf{Cassegrain system:}} \\                         

Equivalent focal length & $f_{eq}$= &96\,m \\
Focal aperture & F\#= &8 \\
Magnification factor & M= &20 \\
Distance primary focus to sub-ref vertex & $\Delta$= &0.29414\,m \\
Distance primary to sub-ref vertex & dvv= &4.50586\,m \\
Distance primary vertex to Cassegrain focus & dvf= &1.377\,m \\
\hline
\end{tabular}
\label{table:LLAMA_optical_parameters}
\end{table}

\subsection{Receivers} 
\label{Receivers}

The receivers to be installed in the Nasmyth cabins of LLAMA are identical to those used in the ALMA radio telescopes. These receivers are planned to be used without any modification, since they have been validated with years of observation, avoiding in this way unnecessary delays during the initial stage of the project, due to the development of new receivers, allowing resources to focus on other subsystems (e.g., the tertiary optical system).   
In the case of ALMA, the observation frequency band of 30-950\,GHz was subdivided into 10 sub-bands, from band 1 to 10, guided by considerations such as atmospheric conditions, astronomical requirements, available detection technology at the time, and polarization considerations \cite{wootten2000frequency, bachiller2000report}. For each sub-band, a dedicated single-pixel heterodyne receiver was integrated into modular cartridges, which are subsequently installed in a cryostat (cryo) at the focal plane of the Cassegrain antenna system \cite{carter2007alma}.

The modular cartridges are divided into two parts: 1) the cold cartridge assembly (CCA) with three cooled stages at 4, 15, and 90\,K installed inside the cryostat, which contains the feed horn, cold optics, and part of the front-end electronics (radio frequency amplifiers, local oscillator, and mixer for the intermediate frequency with its cryogenic amplifiers); 2) the warm cartridge assembly (WCA) at room temperature installed beneath the cryostat, which also contains part of the front-end electronics. Each receiver detects both polarizations, with its feed horn illuminating the sub-reflector through windows on the top of the cryostat. The receivers, equipped with reflective (mirrors) or refractive (dielectric lenses) optics \cite{carter2004alma}, are designed to achieve a 12\,dB edge taper on the sub-reflector to maximize antenna efficiency \cite{lamb2001alma, lamb2003low}. The ten bands observe different parts of the sky, and the observing band in ALMA is selected by tilting the sub-reflector.

\begin{figure}[H]
	\centering
	\includegraphics[width=0.7\textwidth]
    {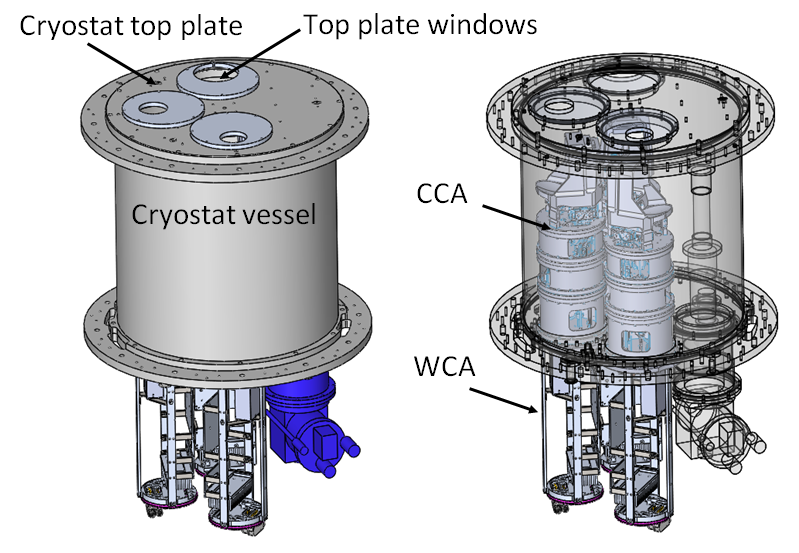}	
	\caption{CAD model of the LLAMA cryostat with cartridges installed.}
		\label{fig:Cartridge and cryostat}
	\end{figure}

For LLAMA, the cartridges will be installed in two cryostats, each with a diameter of 538\,mm and a height of 550\,mm. These cryostats will be placed in each Nasmyth cabin, accommodating three receivers each (Fig. \ref{fig:Cartridge and cryostat}). The focal position ($\omega_{NASS}$) with respect to (wrt) the cryostat top plate for each receiver is shown in Table \ref{table:LLAMA cartridge position}.

\subsection{LLAMA Cabins}
\label{LLAMA cabins}

\begin{figure}[H]
	\centering
	\includegraphics[width=1\textwidth]
    {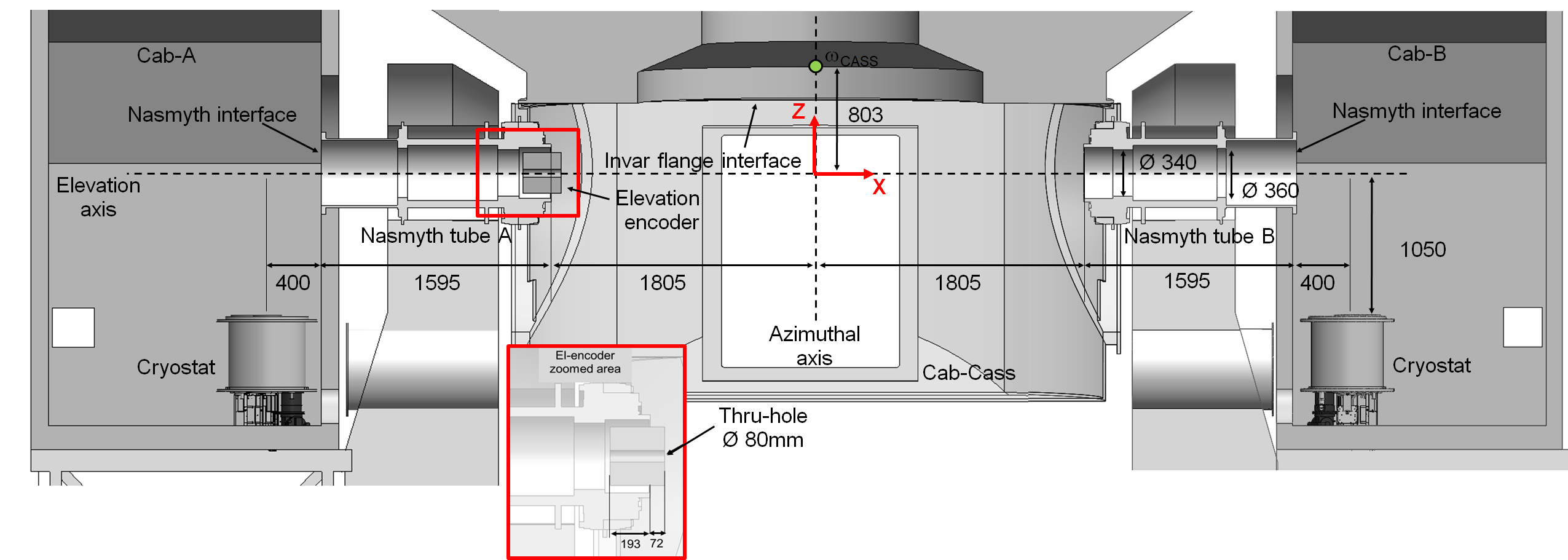}	
	\caption{LLAMA cabins configuration (dimensions in mm).}
		\label{fig:LLAMA_cabins}
	\end{figure}

The LLAMA radio telescope has three cabins for housing signal detection instrumentation: the Cab-Cass, located just below the primary reflector, and the two Nasmyth cabins (Cab-A and Cab-B) on either side. The beam can pass from Cab-Cass to both Nasmyth cabins through the Nasmyth tubes, which have a clearance diameter of 340\,mm for Cab-B and 80\,mm for Cab-A due to the presence of the antenna's elevation encoder. This constraint determined the placement of low-frequency receivers in Cab-B (wider beams) and high-frequency receivers in Cab-A (narrower beams). Each cryostat is located within its respective Nasmyth cabin, with its center positioned 3800\,mm from the Az axis and its reference plane (top cover) 1050\,mm below the El axis of the antenna (Fig. \ref{fig:LLAMA_cabins}). For positioning components inside the antenna, a right-handed rectangular coordinate system (i.e., X, Y, Z) is defined with the origin at the intersection of the azimuthal (Az) and elevation (El) axes.

\begin{figure}[H]
	\centering
	\includegraphics[width=0.8\textwidth]
    {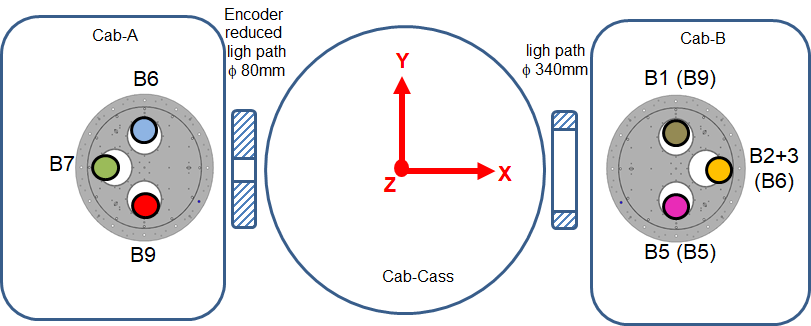}	
	\caption{LLAMA Receivers position scheme. The receiver corresponding to FL phase is indicated in parentheses.}
		\label{fig:LLAMA receivers layout}
	\end{figure}

The position and orientation of each receiver with respect to the LLAMA Coordinate system is presented in Fig. \ref{fig:LLAMA receivers layout}
 and Table \ref{table:LLAMA cartridge position}.

\begin{table}[H]
\caption{LLAMA cartridge position and characteristics for FL and LT configurations.} 
\centering
\begin{tabular}{|ccccccc|}
\hline
\begin{tabular}[c]{@{}c@{}}Band\end{tabular} & \begin{tabular}[c]{@{}c@{}}Freq. \\band \\(GHz)\end{tabular} & \begin{tabular}[c]{@{}c@{}}Cartridge\\ diameter \\(mm)\end{tabular} & \begin{tabular}[c]{@{}c@{}}Cartridge\\ X ; Y \\ position \\ (mm)\end{tabular} & \begin{tabular}[c]{@{}c@{}}Beam\\ X ; Y \\ position \\ (mm)\end{tabular} & \begin{tabular}[c]{@{}c@{}} Angle\\ to\\ cryo center \\ (deg)\end{tabular} & \begin{tabular}[c]{@{}c@{}}Focus\\ position \\ wrt cryo\\ top plate \\(mm)\end{tabular}\\ 
\hline
B1 (LT) &35-52 & 170 & 3800 ; 150 & 3800 ; 190 & 2.48 & 553 \\
B2+3 (LT) &67-116 & 170 & 3950 ; 0 & 3990 ; 0 & 2.48 & 131.8 \\
B5 (FL/LT) &163-211 & 170 & 3800 ; -150 & 3800 ; -200 & 2.38 & -15 \\
B6 (FL) &211-275 & 170 & 3950 ; 0 & 4000 ; 0 & 2.38 & -15 \\
B6 (LT) &211-275 & 170 & -3800 ; 150 & 3800 ; 200 & 2.38 & -15 \\
B7 (LT) &275-373 & 170 & -3950 ; 0 & -4000 ; 0 & 0.97 & -50 \\
B9 (FL) &602-720 & 170 & 3800 ; 150 & 3800 ; 200 & 0.97 & -27 \\
B9 (LT) &602-720 & 170 & -3800 ; -150 & -3800 ; -200 & 0.97 & -27 \\

\hline
\end{tabular}
\label{table:LLAMA cartridge position}
\end{table}

\section{Tertiary Optical System Design}
\label{Tertiary optical system design} 

The tertiary optical system is responsible for coupling the main beam of the telescope, focused onto the antenna focal plane ($\omega_{CASS}$), to the beam of each receiver. For LLAMA, this tertiary optical system is referred to as the NAsmyth Cabins Optical System (NACOS). 

NACOS is expected to maximize the antenna aperture efficiency for all detectors while minimizing optical aberrations. It must be desigend as a flexible system, allowing the antenna user to select between each receiver for single-band observation mode or to enable multi-band simultaneous observations by splitting the sky signal through the implementation of beam splitters (dichroic filters and/or polarizing grids). These upgrades are expected to be part of a future enhancement of the radio telescope and are not analyzed in the current work.

NACOS was designed using quasi-optics (QO) \cite{paul1998goldsmith}, a precise approach for millimeter and submillimeter wavelength systems. At these scales, component sizes, such as mirrors, are comparable to the wavelength of the radio waves, leading to significant diffraction effects that simpler approaches, like geometrical optics (which treats light as rays), cannot accurately model.

In QO, beam propagation is characterized by a Gaussian electric field distribution, known as the fundamental mode. Higher-order Gaussian modes can be superimposed on this fundamental mode to describe more complex electric field distributions.

The propagation of the Gaussian fundamental mode approach has been extensively used in the design of millimeter/sub-millimeter optical systems for astronomical instrumentation \cite{nystrom2009optics, gonzalez2014optics, rioja2010precise, 8440767} and demonstrated \cite{gonzalez2015design} to be a very accurate tool for the design of NACOS for LLAMA.

The propagation of a Gaussian beam in free space is characterized by the following parameters: 

\begin{equation}
R\left(z\right) = z +\frac{1}{z}\left(\frac{\pi\omega_0^2}{\lambda}\right)^2,
\label{Eq:R(z)_2}
\end{equation}

\begin{equation}
\omega\left(z\right) = \omega_0 \left[1+\left(\frac{\lambda z}{\pi\omega_0^2}\right)^2\right]^{0.5},
\label{Eq:w(z)_2}
\end{equation}

\begin{equation}
tan \left(\phi_0\right) = \frac{\lambda z}{\pi\omega_0^2},
\label{Eq:tan_phi0}
\end{equation}
where $R$ is the beam radius of curvature, $\omega$ is the beam transverse radius, $\phi_0$ is the phase shift and $z$ is the distance of propagation of the beam from the beam waist radius $\omega_0$.

The transformation that the Gaussian beam undergoes as it propagates through an optical component or system can be obtained from the ABCD matrix method, where each optical component is described by a particular matrix. The complete system matrix is obtained by multiplying the individual matrices from right to left in the order that the beam encounters them as it propagates\cite{paul1998goldsmith}.

The distance $d_{out}$ of the output beam waist $\omega_{0out}$ to the optical system/component can be found with

\begin{equation}
d_{out}=-\frac{(A d_{in}+B)(C d_{in}+D)+AC{z_c}^2}{(C d_{in}+D)^2+C^2{z_c}^2}
\label{Ec: dout}
\end{equation}
and

\begin{equation}
\omega_{0out}=\frac{\omega_{0in}}{\left[(Cd_{in}+D)^2+C^2{z_c}^2\right]^{0.5}},
\label{Ec: omega out}
\end{equation}
where $d_{in}$ is the distance of the input beam waist $\omega_{0in}$ to the optical system/component. $z_c$=$\pi\omega_0^2/\lambda$ is the confocal distance and $A$, $B$, $C$ and $D$ are elements of the matrix describing either a mirror with focal length $f$ or a free space distance $d$ for NACOS, and are given by

\begin{equation}
M_{mirror}=\begin{bmatrix}
1 & 0 \\
-1/f & 1 \\ 
\end{bmatrix}
;\quad M_{distance}=\begin{bmatrix}
1 & d \\
0 & 1 \\ 
\end{bmatrix}.
\label{Ec: Matrice}
\end{equation}

\begin{figure}[H]
	\centering
	\includegraphics[width=0.8\textwidth]{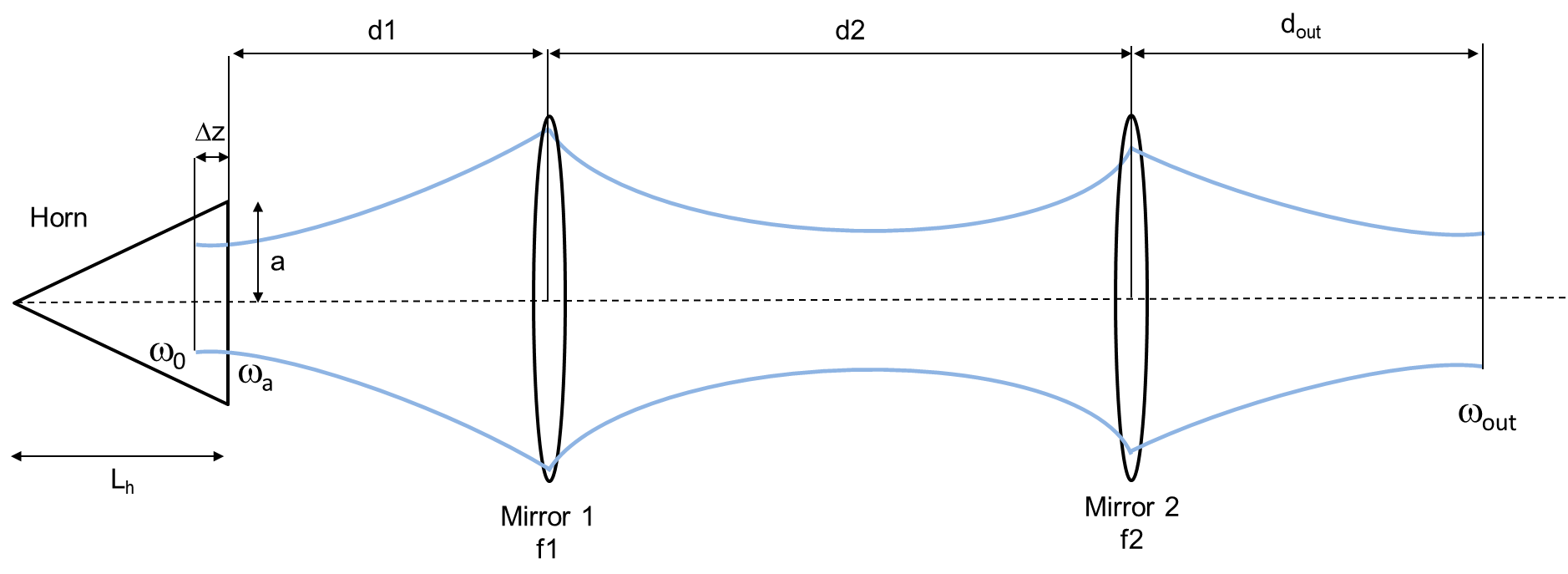}	
	\caption{Conceptual scheme of the quasi-optic system formed by the horn and the optics of the receivers.}
		\label{fig:ALMA_cold_optics_schematic}
	\end{figure}

Having obtained the position and beam waist of the output beam (i.e., $d_{out}$ and $\omega_{0out}$), its subsequent evolution can be determined with Eqs. (\ref{Eq:R(z)_2}) to (\ref{Eq:tan_phi0}). Thus, the Gaussian beam launched from the Nasmyth focus (or from the feed including the cartridge optics, as shown in Fig. \ref{fig:ALMA_cold_optics_schematic} and detailed in Table \ref{table:ALMA_cold_optics_receiver_parameters}), can be propagated through the system. From the size of the beam ($\omega$), it is possible to estimate the losses\cite{murphy1987distortion} due to beam distortion ($K_{Fi}$) [Eq. (\ref{K Fi})] and cross polarization ($K_{COi}$) [Eq. (\ref{K COi})] of each mirror, as well as the defocus loss\cite{paul1998goldsmith} ($K_{Axiall}$) [Eq. (\ref{K axiall})]. Consequently, the total system loss ($K_{Total}$) [Eq. (\ref{K Total})] can be determined. Losses due to spillover on the optical components are not considered at this stage of the analysis.

\begin{equation}
K_{Fi} = 1-\frac{{\omega_{mi}}^2tan^2\theta_i}{8{f_i}^2}.
\label{K Fi}
\end{equation}
\begin{equation}
K_{COi} = 1-\frac{{\omega_{mi}}^2tan^2\theta_i}{4{f_i}^2}.
\label{K COi}
\end{equation}
\begin{equation}
K_{Axiall} = \frac{4}{\left(\frac{\omega_{0b}}{\omega_{0a}}+\frac{\omega_{0a}}{\omega_{0b}}\right)^2+\left(\frac{\lambda\Delta{z}}{\pi\omega_{0a}\omega_{0b}}\right)^2}.
\label{K axiall}
\end{equation}
\begin{equation}
K_{Total} = \left(\prod_{i}K_{Fi}\right)\left(\prod_{i}K_{COi}\right)K_{axiall}.
\label{K Total}
\end{equation}

$\omega_{mi}$, $\theta_i$ and $f_i$ are the beam radius at the mirror $i$, the mirror inclination with respect to the optical axis and the mirror focal length respectively. $\Delta z$ is the distance between the beam waists of beams $a$ and $b$.

Then, from the beam size at the sub-reflector ($\omega_{subref}$), the edge taper on it can be calculated, 

\begin{equation}
Te = 8.696\,\alpha \quad \left[\rm dB\right] \quad ; \quad \alpha=\left(\frac{r_{subref}}{\omega_{subref}}\right)^2
\label{Te}
\end{equation}
and, following the equivalent paraboloid representation, the antenna aperture efficiency\cite{paul1998goldsmith} can be evaluated at the rim of the sub-reflector\cite{pontoppidan2008electromagnetic} 

\begin{equation}
Ap_{eff} = \frac{2}{\alpha}\left(e^{-fb^2\alpha}-e^{-\alpha} \right)^2.
\label{Ap_eff}
\end{equation}

The antenna aperture efficiency considers spillover and illumination over the sub-reflector ($\alpha$) as well as the fractional blockage of the sub-reflector on the primary reflector, given by $fb=r_{subref}/r_p=0.0625$. The maximum achievable aperture efficiency for LLAMA, with fundamental Gaussian beam illumination, is 80.35\%.

\begin{table}[H]
\caption{Quasi-optical parameters for the central frequency of the optics for each receiver.}
\centering
\begin{tabular}{|r|cccccc|}
\hline
Band                                                                                                             & B1 \cite{Tapia2015}            & B2+3\cite{Finger}            & B5 \cite{carter2007alma}          & B6 \cite{carter2007alma}          & B7 \cite{carter2007alma}          & B9 \cite{carter2007alma}          \\ \hline
Freq. (GHz)                                                                                                         & 42.50                   & 91                     & 187                  & 243                  & 324                  & 661                  \\
$a$ {(}mm{)}                                                                                                       & 15.63                  & 10.69                        & 4.50                  & 3.54                 & 3.00                    & 2.54                \\
$L_h$ {(}mm{)}                                                                                                      & 70                     & 43.43                    & 60                   & 46.54                & 45.68                & 15.52                \\
$d1$ {(}mm{)}                                                                                                      & 195.80                 & 91.69                    & 60.05                & 46                   & 38                   & 44.48                \\
$f1$ {(}mm{)}                                                                                                      & 181                    & 91                       & 32.76                & 29.37                & 25.76                & 24.86                \\
$d2$ {(}mm{)}                                                                                                      & -                      & -                        & 140                  & 158                  & 157                  & 95.91                \\
$f2$ {(}mm{)}                                                                                                      & -                      & -                        & 67.19                & 76.61                & 77.07                & 39.41                \\
$pc$ {(}mm{)}                                                                                                      & 5.21                 & 5.50                   & 4.18                & 3.47                & 3.25                & 9.11                \\
$\omega_{0}\, \textrm{(mm)}$                                                                                                      & 9.38                  & 4.80                    & 2.80                & 2.19                & 1.86                & 1.06                \\
$\omega_{a}\, \textrm{(mm)}$                                                                                                      & 9.38                 & 4.80                    & 2.90                & 2.28                & 1.93                & 1.63                \\
$\omega_{out}\, \textrm{(mm)}$                                                                                                    & 42.54                 & 20.70                   & 9.60                & 7.45                & 5.59                & 2.73                \\
$d_{out}$ {(}mm{)}                                                                                                    & 450.4                & 121.96                  & 235.08              & 229.99              & 215.99              & 149.38              \\
 & & & & & &\\
\begin{tabular}[c]{@{}c@{}}Focal plane wrt\\ cryo top plate {(}mm{)}\end{tabular} &553 & 131.8 & -15 & -15 & -50 & -27 \\
\hline
\end{tabular}
\label{table:ALMA_cold_optics_receiver_parameters}
\end{table}

In Table \ref{table:ALMA_cold_optics_receiver_parameters}, both the position of the focus relative to the top plate of the cryostat and its size ($\omega_{out} \equiv \omega_{NASS}$) were obtained from the physical optics simulation of the horn-optics system. In Appendix \ref{Apendix1} the results for B1 and B2+3 are presented.

\subsection{NACOS-FL}
\label{NACOS-FL}

A reduced version of NACOS, called NACOS-FL, is planned for implementation during the FL phase of the project. In NACOS-FL, receivers for bands 5, 6, and 9 will be installed in Cab-B. The optical and mechanical designs were completed from 2016 to 2017 by Jacob Kooi from NASA's Jet Propulsion Laboratory and Astro Electro-Mechanical Engineering, LLC, respectively. NACOS-FL was manufactured in Araraquara, Brazil, from 2018 to 2019 by the company ALFA Ferramentaria. It is currently in the Assembly, Integration and Verification (AIV) stage at the same company.

A frequency independent optical system was design based on a Gaussian Beam Telescope (GBT), which consists of a pair of focusing components separated by the sum of their focal lengths \cite{paul1998goldsmith}. The key feature of the GBT is that both magnification [Eq. (\ref{M})] and output beam waist location [Eq. (\ref{GBT})] are independent of the signal frequency, a useful and desirable feature in systems that operates over broad bandwidths.

\begin{equation}
\mathfrak{M} = \frac{\omega_{0out}}{\omega_{0in}} = \frac{f_2}{f_1}
\label{M}
\end{equation}

\begin{equation}
d_{out} = \frac{f_2}{f_1}\left( f_1+f_2-\frac{f_2}{f_1}d_{in}\right).
\label{GBT}
\end{equation}

\begin{figure}[H]
	\centering
	\includegraphics[width=0.7\textwidth]{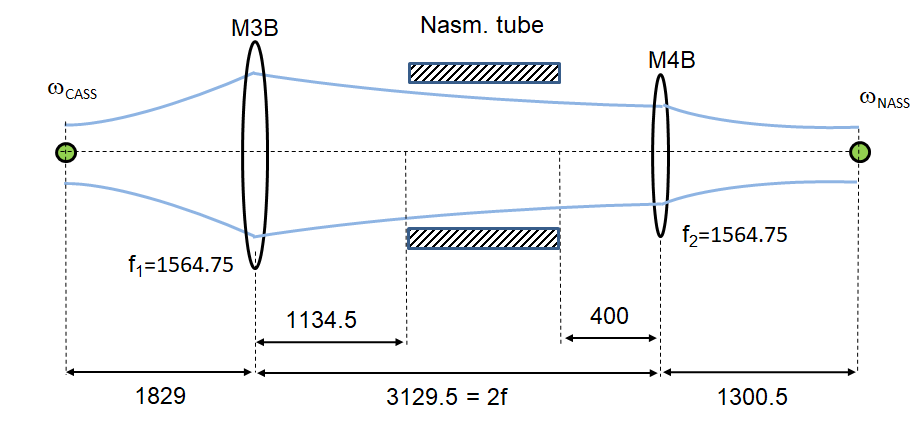}\caption{GBT schematics for NACOS-FL in Cab-B. Units in mm.}
	\label{fig:Gaussian_beam_telescope_cab_B_FL}
\end{figure}

The GBT is formed by mirrors M3B and M4B placed in Cab-CASS and Cab-B respectively, with $\mathfrak{M}$=1 as required by the LLAMA receivers (Fig. \ref{fig:Gaussian_beam_telescope_cab_B_FL}).

\begin{figure}[H]
	\centering
	\includegraphics[width=1.0\textwidth]{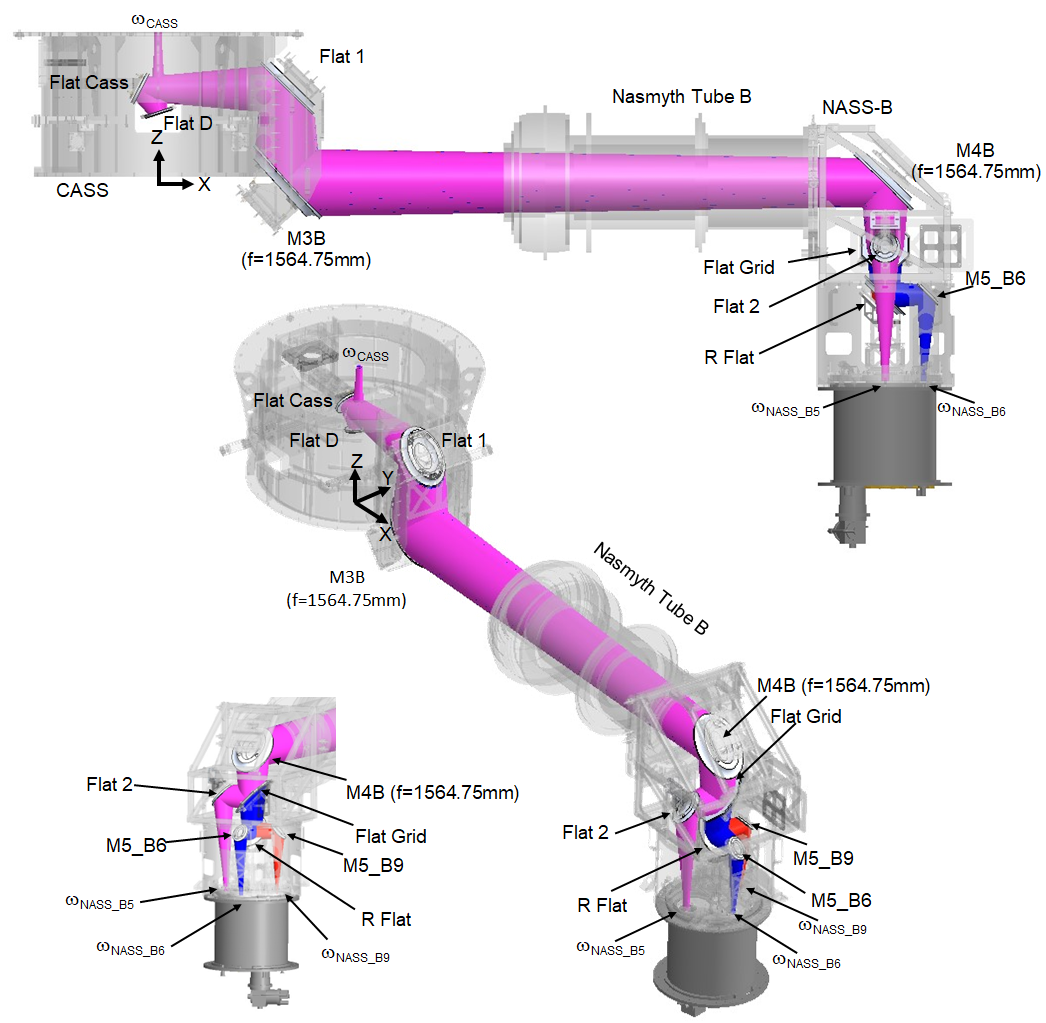}\caption{Configuration of the optical system of NACOS-FL for Cab-B. The Gaussian beam propagation at 30\,dB (99.9\%) contour power is also shown for B5 (purple), B6 (blue) and B9 (read).}
	\label{fig:NACOS FL alambrico_nombres_MULTI}
\end{figure}

The optical system includes a series of flat mirrors. Some mirrors can be removed from the optical path (OP), while others can be reoriented internally using rotary actuators. This flexibility allows for modifications and convenient folding of the beam path. The NACOS-FL 3D model (Fig. \ref{fig:NACOS FL alambrico_nombres_MULTI}) consists of two optomechanical structures: the CASS (in Cab-Cass) and the NASS-B (in Cab-B). 

The beam focused by the Cassegrain system in the antenna's focal plane ($\omega_{CASS}$) is progressively reflected, first, in the common optical components section of the system; the Flat D mirror, the Flat Cass, the Flat 1, the M3B off-axis elliptical mirror and the M4B off-axis hyperbolic mirror. After reflection at M4B, the system's second section allows for the selection of the following optical channels:

\begin{enumerate}
    \item Band 5: By placing the Flat Grid mirror in the OP, the beam is reflected in the plane Y-Z towards Flat 2 mirror, placed in -Y direction, and finally onto the Nasmyth focus of B5 ($\omega_{NASS-B5}$). 

    \item Bands 6 and 9: By removing the Flat Grid from the OP, the beam is allowed to propagate further down until being reflected by R Flat, a rotating mirror that has two set positions in planes X-Z and Y-Z. With R Flat oriented in plane X-Z (+X direction), the beam is reflected towards M5-B6 mirror, and finally onto the Nasmyth focus of B6 ($\omega_{NASS-B6}$). With R Flat oriented in plane Y-Z (+Y direction), the beam is reflected towards M5-B9 mirror, and finally onto the Nasmyth focus of B9 ($\omega_{NASS-B9}$). 
\end{enumerate}

It is worth noting that both the Flat D and Flat Cass mirrors can be removed from the optical path (OP), enabling the signal from the sub-reflector to propagate downstream through the CASS structure.

The main parameters of the mirrors for NACOS-FL configuration are presented in Table \ref{table:FL Mirrors}.

\begin{table}[H]
\caption{Main mirror parameters for FL configuration. Position (X; Y; Z), Surface normal (nx; ny; nz), Diameter ($\varnothing$) and Focal length ($f$).} 
\centering
\scalebox{0.87}{\begin{tabular}{|c|c|ccc|ccc|c|c|}
\hline
Mirror & Band & X(mm) & Y(mm) & Z(mm) & nx & ny & nz & $\varnothing$(mm) & $f$(mm) \\ 
\hline
Flat D &5,6,9 &0 &0 &-423 & -0.3090 &0 &0.9510 &155 & $\infty$  \\
Flat Cass &5,6,9 &-87.22 &0 &-303 &0.4540 &0 &-0.8910 & 176.4 & $\infty$ \\
Flat 1 &5,6,9 &670.50 &0 &-303 &-0.7071 &0 &-0.7071 & 387.6x275 & $\infty$ \\
M3B &5,6,9 &670.50 &0 &-803 &0.7071 &0 &0.7071 & 519.7x369.7 & 1564.75 \\
M4B &5,6,9 &3800 &0 &-803 &-0.7071 &0 &-0.7071 & 425x300 & 1564.75 \\
Flat Grid &5 &3800 &0 &-1153 &0 &-0.7071 &0.7071 & 345 & $\infty$ \\
Flat 2 &5 &3800 &-229.72 &-1153 &0 &0.6923 &-0.7216 & 222x158 & $\infty$ \\
R Flat &6 &3800 &0 &-1403 &0.7071 &0 &0.7071 & 295x209 & $\infty$ \\
M5-B6 &6 &4019.30 &0 &-1403 &-0.6923 &0 &-0.7216 & 155x110 & $\infty$ \\
R Flat &9 &3800 &0 &-1403 &0 &0.7071 &0.7071 & 295x209 & $\infty$ \\
M5-B9 &9 &3800 &208.08 &-1403 &0 &-0.7011 &-0.7131 & 146x103 & $\infty$ \\
\hline
\end{tabular}}
\label{table:FL Mirrors}
\end{table}

\subsection{NACOS-LT}
\label{NACOS-LT}

The Long Term phase of the project involves the installation of receivers in both Nasmyth cabins as detailed in Table \ref{table:LLAMA cartridge position}. In Cab-A, receivers for bands 6 and 9 will be relocated from Cab-B and installed alongside band 7. In Cab-B, receivers for bands 1 and 2+3 will be installed with band 5. Therefore, an upgrade from NACOS-FL to NACOS-LT is necessary. 

\subsubsection{Cab-A}
\label{Cab-A}

A frequency independent optical system was designed for Cab-A. Various alternatives were explored, focusing on reducing system complexity and the number of mirrors to minimize reflection losses and optical loading. This resulted in the development of a three-focusing mirror system, described below.

\begin{figure}[H]
	\centering
	\includegraphics[width=0.59\textwidth]{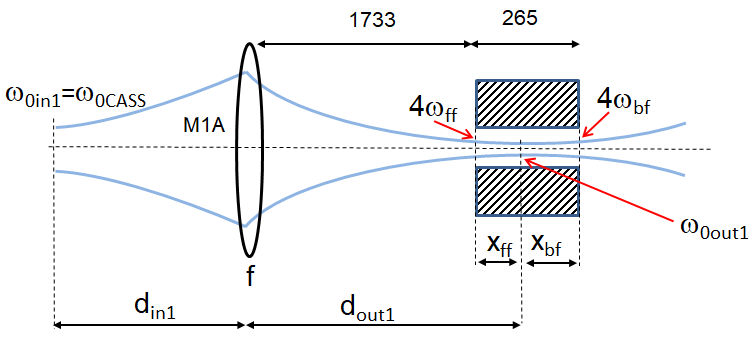}\caption{Gaussian beam configuration to avoid elevation encoder truncation below 4$\omega$. Units in mm.}
	\label{fig:M1A_focal_length}
\end{figure}

The design process began with the determination of the focal length ($f$) of the sole mirror (M1A) to be installed in Cab-Cass. This mirror was positioned with its center coinciding with the intersection of the elevation and azimuth axes, inclined at 45º.
The focal length of M1A ($f_{M1A}$) was optimized to minimize the loss from encoder truncation on the beam. As a design rule, truncation in quasi-optical systems should not occur below 4$\omega$, where $\omega$ is the beam radius in the truncation plane. This ensures a power transmission of $\sim3$5\,dB $\equiv 99.97\%$. This rule was applied to each mirror in the system. Therefore, the output beam waist ($\omega_{out1}$) from M1A must be sized and positioned to ensure that encoder truncation on the beam does not occur below 4$\omega$ on both its front face (ff) and its back face (bf). This is illustrated in Fig. \ref{fig:M1A_focal_length}.

To ensure both $\omega_{ff}$ and $\omega_{bf}$ are simultaneously $<4\varnothing_{enc}$ in Eq. (\ref{Eq:w(z)_2}), and using Eqs. (\ref{Ec: dout}) and (\ref{Ec: omega out}), along with the ABCD elements of a single mirror [Eq. (\ref{Ec: Matrice})], two equations can be derived to determine $f_{M1A}$. Equation (\ref{Ec 1 distancia focal M1A}) addresses truncation at the front face and Eq. (\ref{Ec 2 distancia focal M1A}) addresses truncation at the back face.

\begin{equation}
\left(\frac{\varnothing_{enc}}{4}\right)^2\geq
\frac{{\omega_{0in1}}^2}{\left(1-
\frac{d_{in1}}{f_{ff}}\right)^2+\left(\frac{\pi{\omega_{0in1}}^2}{\lambda{f_{ff}}}\right)^2}
\left[1+
\left(\frac{f_{ff}+\frac{d_{in1-f_{ff}}}{\left(1-
\frac{d_{in1}}{f_{ff}}\right)^2+\left(\frac{\pi{\omega_{0in1}}^2}{\lambda{f_{ff}}}\right)^2}-1733}{\frac{\pi}{\lambda}\frac{{\omega_{0in1}}^2}{\left(1-
\frac{d_{in1}}{f_{ff}}\right)^2+\left(\frac{\pi{\omega_{0in1}}^2}{\lambda{f_{ff}}}\right)^2}}
\right)^2
\right]
\label{Ec 1 distancia focal M1A}
\end{equation}

\begin{equation}
\left(\frac{\varnothing_{enc}}{4}\right)^2\geq
\frac{{\omega_{0in1}}^2}{\left(1-
\frac{d_{in1}}{f_{bf}}\right)^2+\left(\frac{\pi{\omega_{0in1}}^2}{\lambda{f_{bf}}}\right)^2}
\left[1+
\left(\frac{f_{bf}+\frac{d_{in1-f_{bf}}}{\left(1-
\frac{d_{in1}}{f_{bf}}\right)^2+\left(\frac{\pi{\omega_{0in1}}^2}{\lambda{f_{bf}}}\right)^2}-1998}{\frac{\pi}{\lambda}\frac{{\omega_{0in1}}^2}{\left(1-
\frac{d_{in1}}{f_{bf}}\right)^2+\left(\frac{\pi{\omega_{0in1}}^2}{\lambda{f_{bf}}}\right)^2}}
\right)^2
\right]
\label{Ec 2 distancia focal M1A}
\end{equation}

Each of Eqs. (\ref{Ec 1 distancia focal M1A}) and (\ref{Ec 2 distancia focal M1A}) yields two values for the focal length at any given frequency.  If the result is a complex number, the solution lacks physical significance. 

\begin{figure}[H]
\begin{center}
\begin{tabular}{c}
\includegraphics[height=6.5cm]{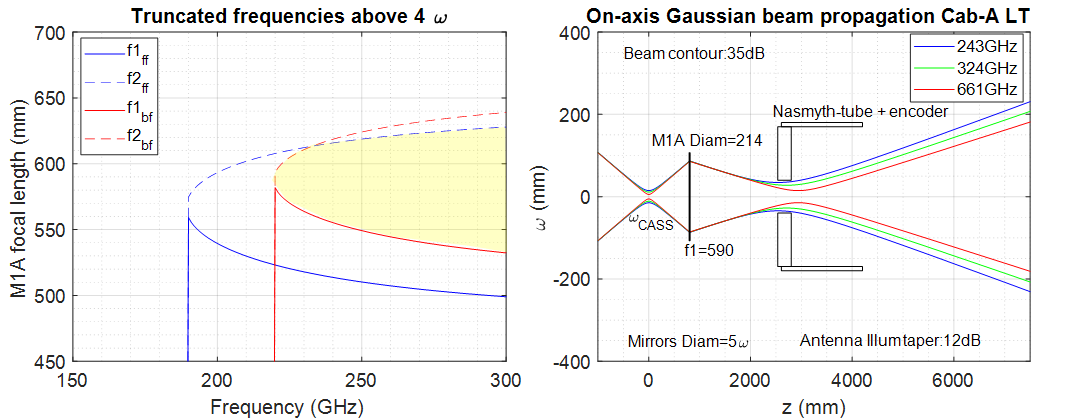}  
\\
(a) \hspace{7.1cm} (b)
\end{tabular}
\end{center}
\caption 
{ \label{fig:Truncated frequencies above 4w and GBP}
(a) Admissible frequencies through encoder with respect to the focal length of M1A mirror. (b) Gaussian beam propagation through encoder of central frequencies of receivers in Cab-A.} 
\end{figure}

Evaluating Eqs. (\ref{Ec 1 distancia focal M1A}) and (\ref{Ec 2 distancia focal M1A}) with $d_{in}$=803\,mm and $\omega_{0in1}\equiv\omega_{CASS}=0.216F\# \lambda Te^{0.5} \sim 6\lambda$, a set of results, shown in Fig. \ref{fig:Truncated frequencies above 4w and GBP} (a) is obtained. The yellow shaded area indicates solutions where truncation below $4\omega$ does not occur on either the front or back face of the encoder. At 220\,GHz, M1A can refocus the beam without truncation below \(4\omega\), with a focal length ranging from 582 to 593\,mm. A focal length of 590\,mm was chosen for M1A, which is expected to produce approximately 0.06\% loss due to encoder truncation at the lowest frequency of B6 (211\,GHz), a value considered negligible. Fig. \ref{fig:Truncated frequencies above 4w and GBP} (b) shows the propagation of the Gaussian beams through the encoder.

Since $\mathfrak{M}=1$ for NACOS, Eqs. (\ref{M}) and (\ref{GBT}) show that the output beam waist would be 1577\,mm from the center of M1A, still inside Cab-Cass (Fig. \ref{fig:LLAMA_cabins}). Therefore, completing the frequency independent system for Cab-A requires the addition of at least two more focusing mirrors.

\begin{figure}[H]
	\centering
	\includegraphics[width=0.8\textwidth]{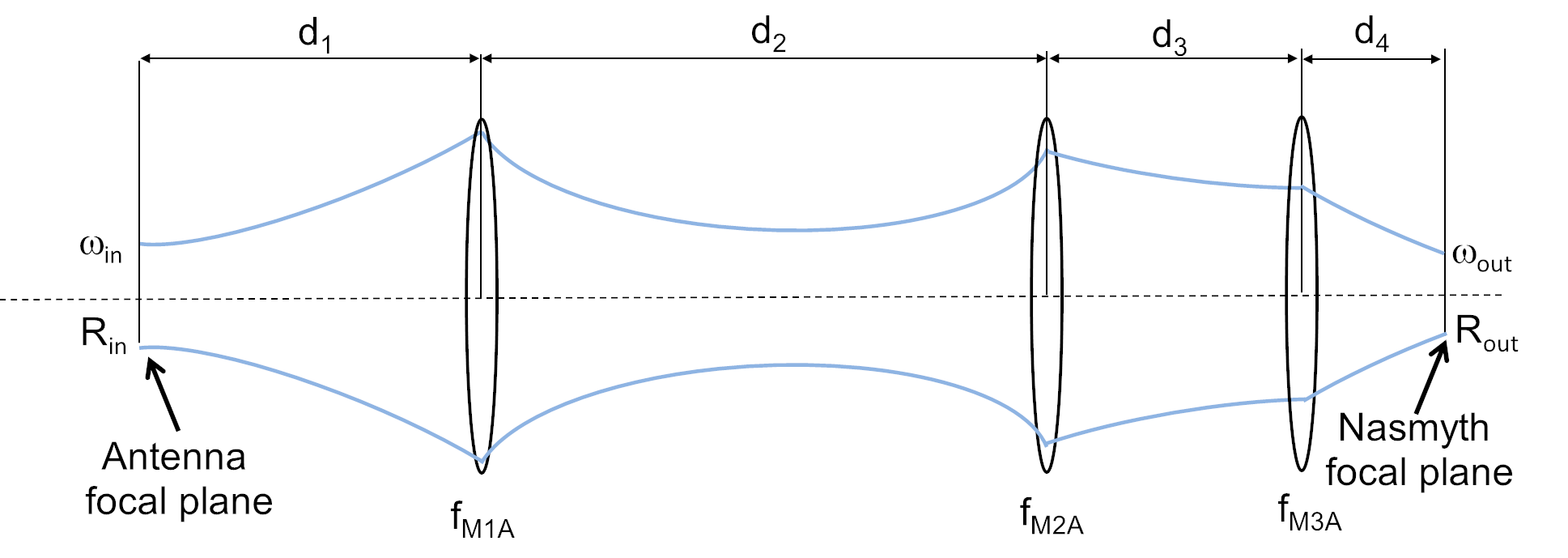}
    \caption{Schematic of the quasi-optical system to be implemented in Cab-A for LT. 
    $f_{M1A}$, $f_{M2A}$, and $f_{M3A}$ are the focal lengths of mirrors M1A, M2A, and M3A respectively.
    $d_1$ and $d_4$ are the distances between mirrors and focal planes. $d_2$ and $d_3$ are distances between mirrors.}
	\label{fig:Freq_indep_three_mirror_schematic}
\end{figure}

The methodology for the implementation of a frequency independent system formed by N-focusing mirrors is well described in \cite{gonzalez2016frequency}. In an optical system formed by N components (Fig. \ref{fig:Freq_indep_three_mirror_schematic}), the elements of the ABCD matrix describing the entire system ($M_N$) can be described in terms of the elements of the $M_{N-1}$ matrix. For the current configuration of three mirrors (N = 3), $M_2$ is the system formed by mirrors M1A and M2A. A new focusing component with focal length $f_3$ (i.e., mirror M3A) and propagation distance to the new target plane $d_{N+1}$ (i.e., $d_4$) is added. The following equations allow the design of the system: 

\begin{equation}
C_{3}=C_{2}-\frac{A_{2}}{f_3}
\label{Ec: 37 Gonzalez "caso actual"}
\end{equation}
\begin{equation}
\frac{1}{f_3}=\frac{1}{d_{4}}+\frac{D_{2}}{B_{2}}
\label{Ec: 44 Gonzalez "caso actual"}
\end{equation}
\begin{equation}
\omega_{out}=\frac{d_{4}}{|B_{2}|}\omega_{in}
\label{Ec: 45 Gonzalez "caso actual"}
\end{equation}
\begin{equation}
\frac{1}{R_{out}}=\frac{1}{d_{4}}\left[1+\frac{B_{2}}{d_{4}}\left(A_{2}+\frac{B_{2}}{R_{in}}\right)\right]
\label{Ec: 46 Gonzalez "caso actual"}
\end{equation}

$A$, $B$, $C$ and $D$ are the elements of the matrix that describes the system. The sub-index corresponds to the optical component number. 

The NACOS configuration requires that frequency independence occurs between both focal planes (i.e., $\omega_{CASS}$ and $\omega_{NASS}$). This implies that $R_{in}$ and $R_{out}$ are both infinite.  Introducing these constraints in Eq. (\ref{Ec: 46 Gonzalez "caso actual"}) along with Eq. (\ref{Ec: 45 Gonzalez "caso actual"}) yields:
\begin{equation}
A_2=1.
\label{Ec: A2 3GBT}
\end{equation}

Since the beam converges towards the output beam waist, the value of $B_2$ is negative. This implies that $|B_2|$ is equal to $-B_2$ and introducing this condition in Eq. (\ref{Ec: 45 Gonzalez "caso actual"}), the value of $B_2$ can be calculated as

\begin{equation}
B_2=-d_4.
\label{Ec: d4_3 3GBT}
\end{equation}

On the other hand, the magnification of the entire system ($\mathfrak{M_{3}}$) must be equal to one, which implies that the element $C_3$ of $M_3$ is equal to zero, i.e. an afocal system. This condition, combined with Eq. (\ref{Ec: A2 3GBT}), leads to the following relation:
\begin{equation}
C_2=\frac{1}{f_{M3A}}.
\label{Ec: f3 3GBT}
\end{equation}

The matrix $M_2$ is defined as,

\begin{equation*}
M_2=M_{d3} M_{M2A} M_{d2} M_{M1A} M_{d1}
=\begin{bmatrix}
1 & d_3 \\
0 & 1 \\
\end{bmatrix}
\begin{bmatrix}
1 & 0 \\
-\frac{1}{f_{M2A}} & 1 \\
\end{bmatrix}
\begin{bmatrix}
1 & d_2 \\
0 & 1 \\
\end{bmatrix}
\begin{bmatrix}
1 & 0 \\
-\frac{1}{f_{M1A}} & 1 \\
\end{bmatrix}
\begin{bmatrix}
1 & d_1 \\
0 & 1 \\
\end{bmatrix},
\end{equation*}

in which replacing $d_1=803\rm\,mm$ (distance from Cassegrain focal plane to center of mirror M1A), $d_2=3800\rm\,mm$ (distance between the centers of mirrors M1A and M2A) and $f_{M1A}=590\rm\,mm$ (focal length of mirror M1A) gives

\begin{equation}
M_2=\begin{bmatrix}
5.441\left(\frac{d_3}{f_{M2A}}\right)-0.0017d_3-5.441 & 568.86\left(\frac{d_3}{f_{M2A}}\right)-0.361d_3-568.86  \\
5.441\left(\frac{1}{f_{M2A}}\right)-0.0017 & 568.86\left(\frac{1}{f_{M2A}}\right)-0.361 \\
\end{bmatrix}.
\label{Ec: M_2 3GBT}
\end{equation}

\begin{figure}[H]
	\centering
	\includegraphics[width=0.72\textwidth]{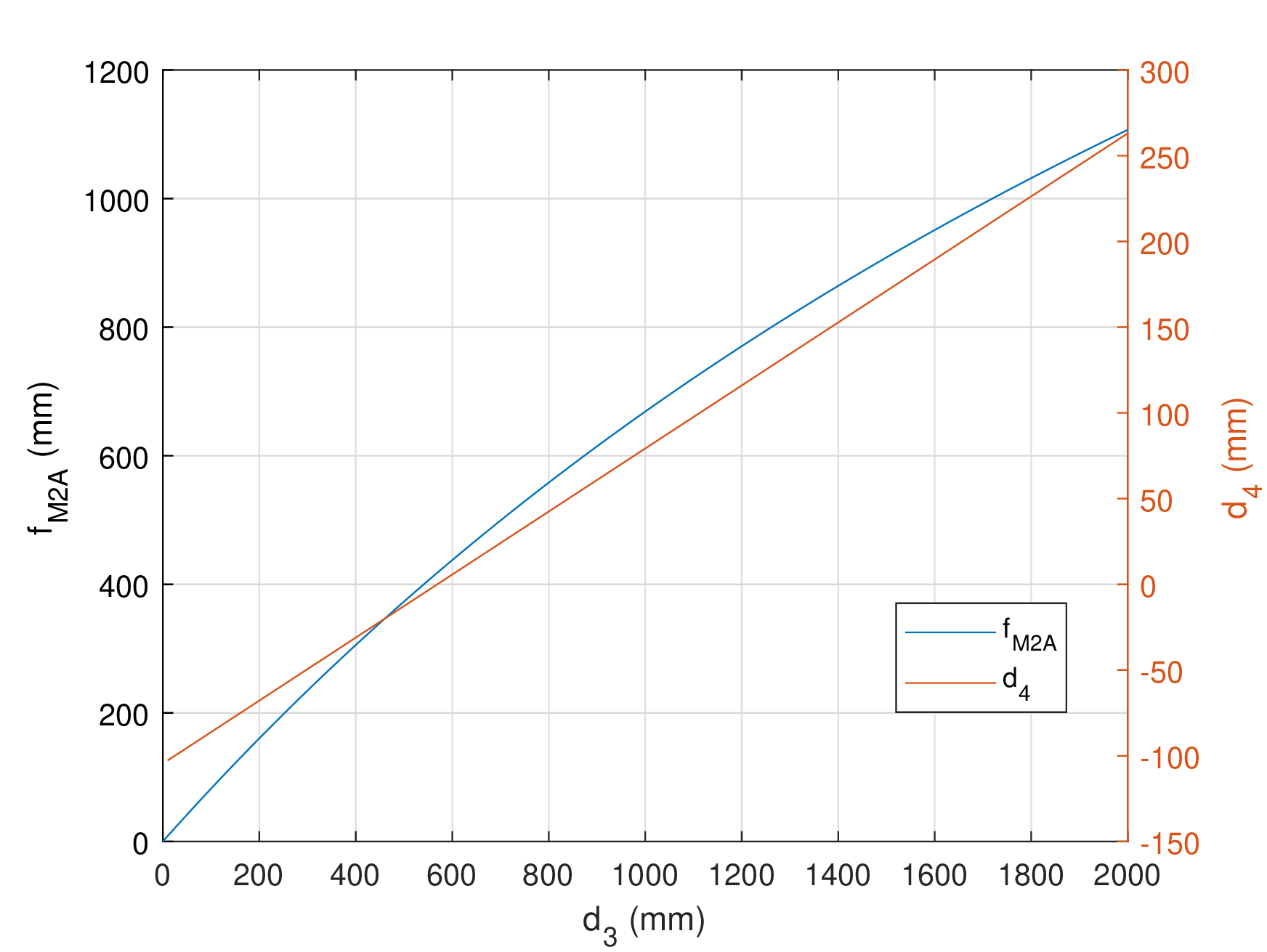}
    \caption{Relationship for the quasi-optical design of Cab-A for LT system: $d_3$ and $f_{M2A}$ (left axis), $d_3$ and $d_4$ (right axis).}
	\label{fig:Cab_A_3GBT_d3_vs_f2_d4}
\end{figure}

Combining Eqs. (\ref{Ec: A2 3GBT}), (\ref{Ec: d4_3 3GBT}) and (\ref{Ec: f3 3GBT}) with the elements of $M_2$ [Eq. (\ref{Ec: M_2 3GBT})], the design Eqs. (\ref{Ec: Final eq 1 3GBT}), (\ref{Ec: Final eq 2 3GBT}) (also plotted in Fig. \ref{fig:Cab_A_3GBT_d3_vs_f2_d4}) and (\ref{Ec: Final eq 3 3GBT}) for the system are found:

\begin{equation}
5.441\left(\frac{d_3}{f_{M2A}}\right)-0.0017d_3-5.441=1
\label{Ec: Final eq 1 3GBT}
\end{equation}

\begin{equation}
568.86\left(\frac{d_3}{f_{M2A}}\right)-0.361d_3-568.86=-d_4
\label{Ec: Final eq 2 3GBT}
\end{equation}

\begin{equation}
5.44\left(\frac{1}{f_{M2A}}\right)-0.0017=\frac{1}{f_{M3A}}.
\label{Ec: Final eq 3 3GBT}
\end{equation}

From Fig. \ref{fig:Cab_A_3GBT_d3_vs_f2_d4} it is seen that $d_3\sim 5.5d_4+570\rm\,mm$. This means that $d_3$ increases faster than $d_4$. Therefore $d_4=150\rm\,mm$ is established as a good compromise between the location of the mirror M3A outside the cryostat and the distance between mirrors M2A and M3A ($ d_3=1385\rm\,mm$). The set of parameters that completes the optical characteristics of the system for Cab-A are:

\begin{table}[H]
\caption{Final optical parameters of Cab-A LT system.} 
\centering
\scalebox{0.87}{\begin{tabular}{|c|c|}
\hline
Parameter & Value (mm) \\ 
\hline
$f_{M1A}$ &590  \\
$f_{M2A}$ &857.4 \\
$f_{M3A}$ &215 \\
$d_1$ &803 \\
$d_2$ &3800 \\
$d_3$ &1385 \\
$d_4$ &150 \\

\hline
\end{tabular}}
\label{table:Cab-A LT optical parameters}
\end{table}

\begin{figure}[H]
	\centering
	\includegraphics[width=0.75\textwidth]{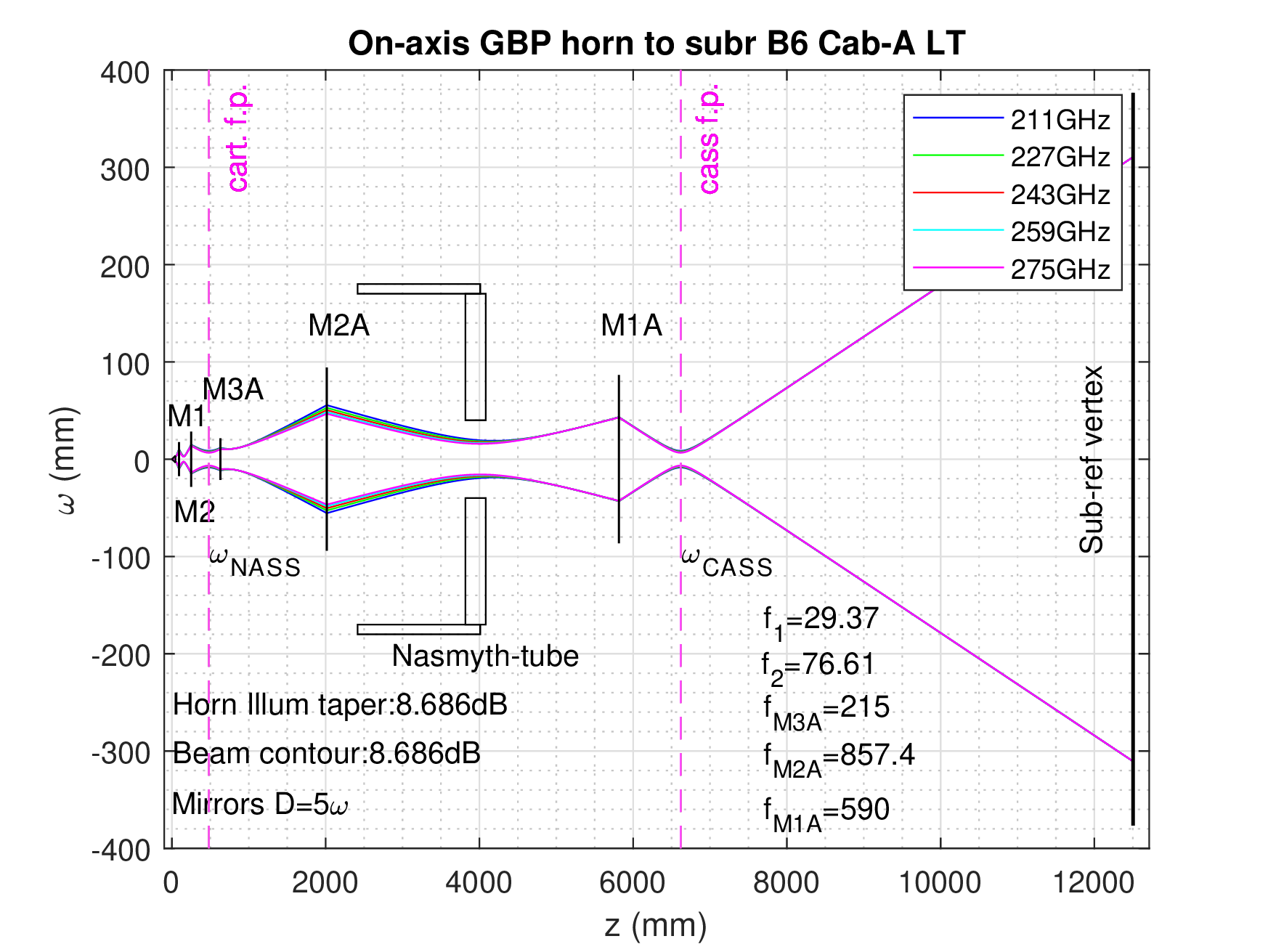}\caption{Gaussian beam propagation from the B6 receiver's feed horn up to the sub-reflector, for the Cab-A optical system configuration.}
	\label{fig:Cab-A_B6_3GBT_GBP_horn_to_antenna}
\end{figure}

The QO propagation of the fundamental Gaussian beam from the feed horn, along the optical system up to the sub-reflector, was performed for five spaced frequencies on each band. ( e.g., B6 for Cab-A in Fig. \ref{fig:Cab-A_B6_3GBT_GBP_horn_to_antenna}). From this analysis, the size and radii of curvature for the focusing mirrors were obtained, as well as the evaluation of the system optical performance [Eqs. (\ref{K Fi}) to (\ref{Ap_eff})]. Refer to Appendix \ref{Apendix2} for details.

\begin{figure}[H]
	\centering
	\includegraphics[width=1.0\textwidth]{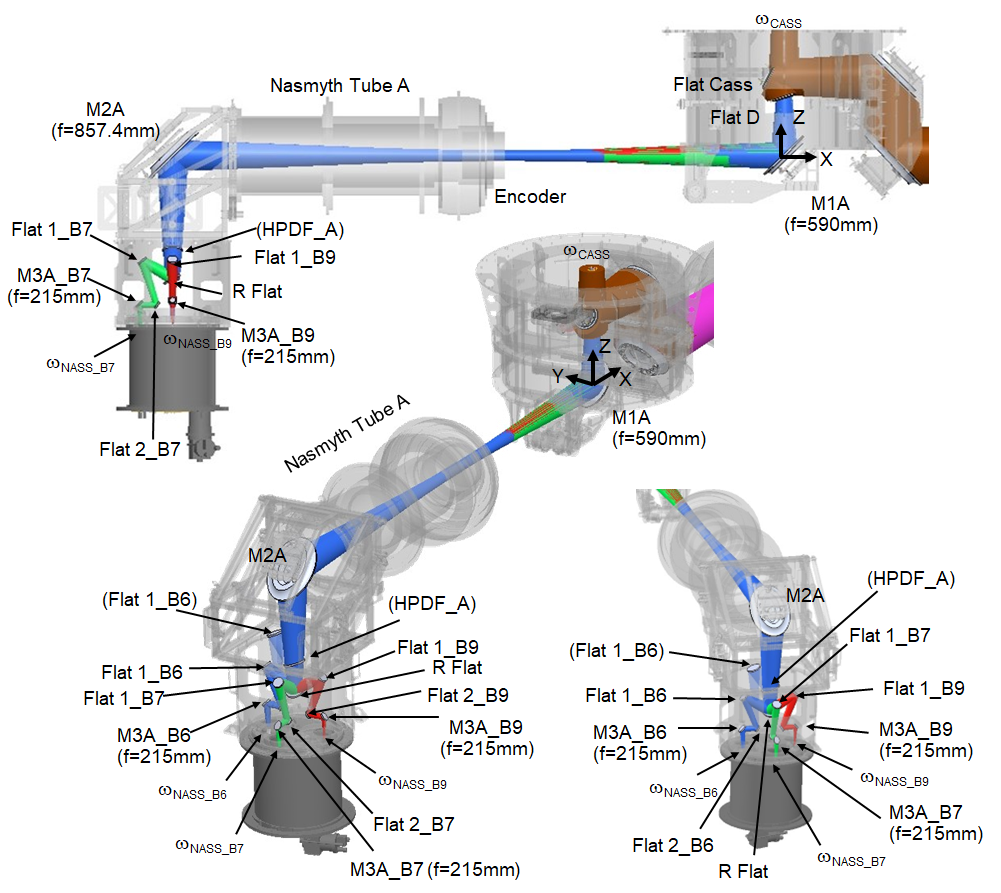}\caption{Configuration of the optical system of NACOS-LT for Cab-A. The Gaussian beam propagation at 35\,dB (99.96\%) contour power is also shown for B6 (blue), B7 (green) and B9 (read).}
	\label{fig:NACOS LT A alambrico_nombres_MULTI}
\end{figure}

To accommodate the optical system, a series of flat mirrors, some removable and others reorientable with motors, completes the design. This setup allows the beams to be directed such that the same support structure used in Cab-B (i.e., NASS-B)  can be reused in Cab-A as NASS-A, eliminating the need for a new structure in Cab-A
The 3D model of NACOS-LT for Cab-A is shown in Fig. \ref{fig:NACOS LT A alambrico_nombres_MULTI}. consists of two optomechanical structures; the CASS (common for both Nasmyth cabins) and the NASS-A (in Cab-A).

The beam focused by the Cassegrain system in the antenna's focal plane ($\omega_{CASS}$) propagates downstream towards mirror M1A. This is possible either by remotely removing mirror flat D from the optical path (OP) or by replacing it with a high-pass dichroic filter (HPDF) in a future upgrade to allow transmission of higher frequencies (i.e., above B5). The beam is progressively reflected and refocused, first by the off-axis elliptical mirror M1A, which refocuses the beam through the elevation encoder and the Nasmyth Tube A towards Cab-A. The beam is then reflected by the M2A off-axis elliptical mirror, and finally by R Flat, a rotating mirror that allows the selection of the any of the following optical channels in Cab-A:

\begin{enumerate}
    \item Band 6: With R Flat oriented in plane Y-Z (+Y direction), the beam is reflected towards Flat 1-B6, Flat 2-B6, M3A-B6 off-axis elliptical mirror, and finally onto the Nasmyth focus of B6 ($\omega_{NASS-B6}$).
    
     \item Band 7: With R Flat oriented in plane X-Z (-X direction), the beam is reflected towards Flat 1-B7, Flat 2-B7, M3A-B7 off-axis elliptical mirror, and finally onto the Nasmyth focus of B7 ($\omega_{NASS-B7}$).

    \item Band 9: With R Flat oriented in plane Y-Z (-Y direction), the beam is reflected towards Flat 1-B9, Flat 2-B9, M3A-B9 off-axis hyperbolic mirror, and finally onto the Nasmyth focus of B9 ($\omega_{NASS-B9}$).
\end{enumerate}

Two important features of the system are:
\begin{itemize}
    \item The M1A mirror is supported by a rotating system that allows its complete removal from the optical path. Together with the removal of mirrors Flat D and Flat Cass, this setup clears the path, enabling the signal to propagate inside Cab-Cass for additional receivers, cameras, or optical systems that may be installed in this cabin.
    \item In a future upgrade, a high-pass dichroic filter (HPDF) could be installed between mirrors M2A and R Flat. This filter would reflect B6 while transmitting B7 and B9, enabling simultaneous observations in two bands in Cab-A. The analysis and design of the HPDF are beyond the scope of this work.
\end{itemize}

The main parameters of the mirrors for the LT configuration are presented in Table \ref{table:LT Cab-A Mirrors}.

\begin{table}[H]
\caption{Mirror parameters for the LT configuration in Cab-A. Position (X; Y; Z), Surface normal (nx; ny; nz), Diameter ($\varnothing$) and Focal length ($f$).} 
\centering
\scalebox{0.87}{\begin{tabular}{|c|c|ccc|ccc|c|c|}
\hline
Mirror & Band & X(mm) & Y(mm) & Z(mm) & nx & ny & nz & $\varnothing$(mm) & $f$(mm) \\ 
\hline
M1A &6,7,9 &0 &0 &-803 & -0.7071 &0 &0.7071 &277x178 & 590  \\
M2A &6,7,9 &-3800 &0 &-803 &0.7071 &0 &-0.7071 & 329x223 & 857.4 \\
R Flat &6 &-3800 &0 &-1583 &0 &0.5000 &0.8660 & 115 & $\infty$ \\
Flat 1-B6 &6 &-3800 &199.06 &-1468.09 &0 &-0.6534 &-0.7570 & 67 & $\infty$ \\
Flat 2-B6 &6 &-3800 &100.02 &-1718.15 &0 &0.8271 &0.5620 &52 & $\infty$ \\
M3A-B6 &6 &-3800 &206.25 &-1718.15 &0 &-0.6923 &-0.7216 & 68x47 & 215 \\
R Flat &7 &-3800 &0 &-1583 &-0.5000 &0 &0.8660 & 115 & $\infty$ \\
Flat 1-B7 &7 &-3984.69 &0 &-1476.38 &-0.6227 &0 &-0.7825 & 68 & $\infty$ \\
Flat 2-B7 &7 &-3900 &0 &-1753.02 &0.8039 &0 &0.5948 & 52 & $\infty$ \\
M3A-B7 &7 &-4002.55 &0 &-1753.02 &0.7011 &0 &-0.7131 & 60x42 & 215 \\
R Flat &9 &-3800 &0 &-1583 &0 &-0.5000 &0.8660 & 115 & $\infty$ \\
Flat 1-B9 &9 &-3800 &-195.41 &-1470.18 &0 &0.6439 &-0.7651 & 62 & $\infty$ \\
Flat 2-B9 &9 &-3800 &-100 &-1730.02 &0 &-0.82 &0.5724 & 49 & $\infty$ \\
M3A-B9 &9 &-3800 &-202.53 &-1730.02 &0 &0.7011 &-0.7131 & 50x35 & 215 \\
\hline
\end{tabular}}
\label{table:LT Cab-A Mirrors}
\end{table}

\subsubsection{Cab-B}
\label{Cab-B}

For the LT configuration of the optical system in Cab-B, bands 9 and 6 of FL are replaced by bands 1 and 2+3 respectively. In the case of B5, the OP for LT remains the same as for FL. 

Bands 1 and 2+3 were integrated into the system by reusing as many optical components as possible from the FL version. Some flat mirrors were replaced or added to provide the required beam path length. The reference beam waist $\omega_{NASS}$ for the design was based on the central frequency of each band (Appendix \ref{Apendix1}). It was placed at 1300.5\,mm from M4B for B1 and 1299\,mm for B2+3. the optimal distance for a perfect beam radius match on the sub-reflector should have been 1188\,mm for B1 and 1272\,mm for B2+3. The defocus losses ($\sim 16\lambda_{B1}$ and $\sim 9\lambda_{B2+3}$) are negligible compared to the truncation losses from the existing system (mirrors and Nasmyth Tube B) when placing the $\omega_{NASS}$ in the ideal position.

\begin{figure}[H]
	\centering
	\includegraphics[width=1.0\textwidth]{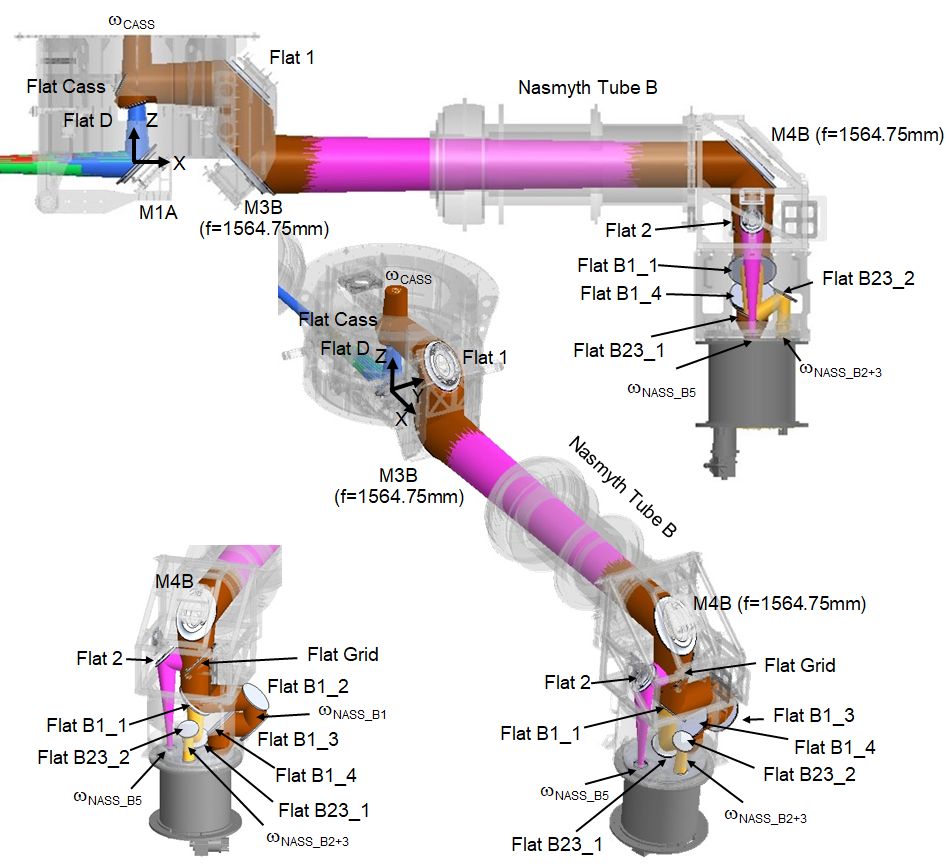}\caption{Configuration of the optical system of NACOS-LT for Cab-B. The Gaussian beam propagation at 30\,dB contour power is also shown for B1 (brown), B2+3 (orange) and B5 (purple).}
	\label{fig:NACOS LT B alambrico_nombres_MULTI}
\end{figure}

The 3D model of NACOS-LT for Cab-B is shown in Fig. \ref{fig:NACOS LT B alambrico_nombres_MULTI}. 

After propagating through the common section of all system bands (Sec. \ref{NACOS-FL}), the system allows the selection of the following optical channels:

\begin{enumerate}
    \item Band 5: The optical system is exactly the same used for NACOS-FL (Sec. \ref{NACOS-FL}).

    \item Band 1: Removing the Flat Grid from the OP allows the beam to propagate further down, where it is progressively reflected (in the Y-Z plane) by four flat mirrors, named flat B1-1 to Flat B1-4. The signal is then focused by a lens into the feed inside the cryostat. The Nasmyth focus of B1 ($\omega_{NASS-B1}$) at the central frequency is located between mirrors Flat B1-2 and Flat B1-3.
    
    \item Band 2+3: By removing the Flat Grid (used for B5) and the Flat B1-1 (used for B1) from the OP, the beam propagates further down. It is progressively reflected (in the plane X-Z, towards +X direction) by two flat mirrors, Flat B23-1 first and Flat B23-2 second. The signal is then focused by the lens into the feed inside the cryostat. The Nasmyth focus of B2+3 ($\omega_{NASS-B2+3}$) of the central frequency is located between mirrors Flat B23-1 and Flat B23-2. 
\end{enumerate}

Both Flat D and Flat Grid are metallic mirrors that could be replaced in an upgrade phase of the LT configuration. Flat D could be replaced by a HPDF, reflecting lower frequencies to Cab-B while transmitting higher frequencies (above B5) to Cab-A. Flat Grid could be replaced by a polarizing grid that reflects one polarization state and transmits the orthogonal one. The analysis of the HPDF and the polarizing grid is beyond the scope of this work.

The parameters of the mirrors for NACOS-LT configuration for Cab-A are presented in Table \ref{table:LT Cab-B Mirrors}.

\begin{table}[H]
\caption{Mirror parameters for LT in Cab-B configuration. Position (X; Y; Z), Surface normal (nx; ny; nz), Diameter ($\varnothing$) and Focal length ($f$).} 
\centering
\scalebox{0.87}{\begin{tabular}{|c|c|ccc|ccc|c|c|}
\hline
Mirror & Band & X(mm) & Y(mm) & Z(mm) & nx & ny & nz & $\varnothing$(mm) & $f$(mm) \\ 
\hline
Flat D &1,2+3,5 &0 &0 &-423 & -0.3090 &0 &0.9510 &155 & $\infty$  \\
Flat Cass &1,2+3,5 &-87.22 &0 &-303 &0.4540 &0 &-0.8910 & 176.4 & $\infty$ \\
Flat 1 &1,2+3,5 &670.50 &0 &-303 &-0.7071 &0 &-0.7071 & 387.6x275 & $\infty$ \\
M3B &1,2+3,5 &670.50 &0 &-803 &0.7071 &0 &0.7071 & 519.7x369.7 & 1564.75 \\
M4B &1,2+3,5 &3800 &0 &-803 &-0.7071 &0 &-0.7071 & 425x300 & 1564.75 \\
Flat Grid &5 &3800 &0 &-1153 &0 &-0.7071 &0.7071 & 345 & $\infty$ \\
Flat 2 &5 &3800 &-229.72 &-1153 &0 &0.6923 &-0.7216 & 222x158 & $\infty$ \\
Flat B1-1 &1 &3800 &0 &-1453 &0 &0.6560 &0.7547 & 320 & $\infty$ \\
Flat B1-2 &1 &3800 &457.39 &-1390.72 &0 &-0.6560 &-0.7547 & 250 & $\infty$ \\
Flat B1-3 &1 &3800 &457.39 &-1613 &0 &-0.7071 &0.7071 & 280 & $\infty$ \\
Flat B1-4 &1 &3800 &200.39 &-1613 &0 &0.6916 &-0.7222 & 280 & $\infty$ \\
Flat B23-1 &2+3 &3800 &0 &-1753 &0.4656 &0 &0.8850 & 195 & $\infty$ \\
Flat B23-2 &2+3 &3800 &0 &-1615.34 &-0.1390 &0 &-0.9903 & 160 & $\infty$ \\
\hline
\end{tabular}}
\label{table:LT Cab-B Mirrors}
\end{table}

\begin{figure}[H]
	\centering
	\includegraphics[width=0.75\textwidth]{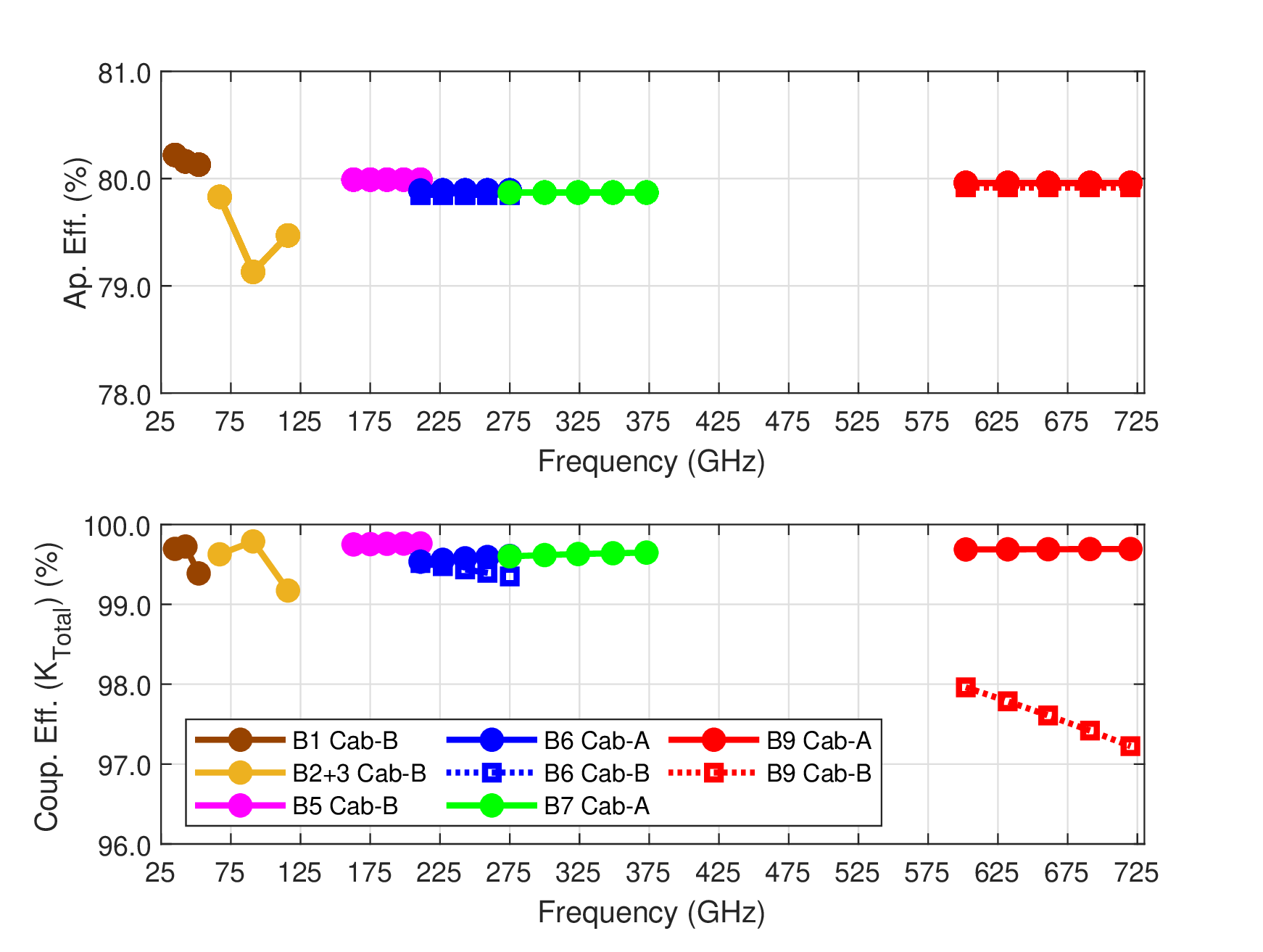}\caption{QO results of aperture efficiency (top) and beam coupling (bottom) for the whole system.}
	\label{fig:QO_results_FL_and_LT}
\end{figure}

The QO Aperture efficiency and beam coupling analysis for all the bands of the system are summarized in Fig. \ref{fig:QO_results_FL_and_LT}. The aperture efficiency plot shows that B1 and B2+3 do not exhibit fully frequency-independent behavior, meaning efficiency does not remain constant across the frequency range. This is due to variations in the receiver focus position ($\omega_{NASS}$) with the frequency (Appendix \ref{Apendix1}). For B2+3, efficiency decreases because the size of $\omega_{NASS}$, causes under-illumination of the sub-reflector, resulting in an edge taper value higher than the expected 12\,dB (Appendix \ref{Apendix1}). The coupling efficiency plot indicates that B9 for Cab-B has a total coupling efficiency $K_{total}$ of $\sim 2.5\%$ less than the others. This is due to the location of $\omega_{NASS-B9}$, which, for unknown reasons to the author of the present work, was placed at 1285.15\,mm (i.e., $\sim 34\lambda_{661\rm\,GHz}$) from M4B instead of the ideal position of 1300.50\,mm.

\section{Physical Optics Simulations}
\label{Physical Optics simulations}

The optical performance of the system for each receiver (presented in Sec. \ref{Tertiary optical system design}), was analyzed through simulations carried out with GRASP \cite{ticraweb}, a dedicated software package for reflector systems widely used in the design, verification and optimization of advanced reflector antenna systems. GRASP uses the Physical Optics (PO) algorithm, which enables the prediction and analysis of the entire far-field and near-field patterns of the system.

In this work, two sets of simulations were performed for the lower, central, and higher frequencies of each receiver channel. The first set of simulations is based on propagating the fundamental Gaussian beam from $\omega_{CASS}$  to an evaluation plane located at $\omega_{NASS}$, passing through the entire tertiary optical system. This set of simulations is identified as reception mode. The second set of simulations is based on propagating the fundamental Gaussian beam from $\omega_{NASS}$ to the sky, passing through the entire tertiary optical system, the Nasmyth tubes, encoder and the Cassegrain system. This set of simulations is identified as transmission mode. In each set of simulations, a series of merit figures were evaluated.

Reception mode merit figures:
 \begin{itemize}
    \item Beam coupling, which is a measure of how much of the energy contained in the beam produced by the system ($E_{out}$) is coupled with a fundamental Gaussian beam ($E_{Gauss}$) in the evaluation plane. It can be obtained by
    
    \begin{equation}
    \eta_{coupling} = \frac{{\left|\iint{E_{out}E_{Gauss}^*dxdy}\right|}^2}{\iint{E_{out}E_{out}^*dxdy}\iint{E_{Gauss}E_{Gauss}^*dxdy}},
    \label{Eq:Gaussicity_orig}
    \end{equation}

    where $^*$ means the complex conjugate. 

    \item Gaussicity \cite{johansson1995comparison}, which is a measure of how well the beam produced by the system ($E_{out}$) maintains its Gaussian shape, indicates the aberration level in the beam. To obtain the Gaussicity, Eq. (\ref{Eq:Gaussicity_orig}) can be used by replacing $E_{Gauss}$ with the fitted Gaussian distribution of $E_{out}$, which can be obtained by

    \begin{equation}
    E_{Gauss-fit} = e^{-\left[\frac{\left(x-x_{offset}\right)^2}{{\omega_{0x}}^2}+\frac{\left(y-y_{offset}\right)^2}{{\omega_{0y}}^2}\right]}.
    \label{Eq:Gauss_fit}
    \end{equation}
    
    \item Ellipticity, which indicates how much the beam deviates from the ideal circular shape of the fundamental Gaussian beam mode, can be determined by 

    \begin{equation}
    Ellipticity = 1-\frac{\min(\omega_{0x}, \omega_{0y})}{\max(\omega_{0x}, \omega_{0y})},
    \label{Eq:Ellipticity}
    \end{equation}

    where $\omega_{0x}$ and $\omega_{0y}$ are the beam waist radius obtained from the Gaussian fit [Eq. (\ref{Eq:Gauss_fit})].
    
    \item Cross-polarization level induced by the system.
\end{itemize}

Transmission mode merit figures:

\begin{itemize}
    \item Antenna aperture efficiency.
    \item Antenna beam pattern on the sky, from which the Full Width to Half Maximum (FWHM), first side lobe level (left and right) and pointing (in Azimuth and Elevation coordinates) were obtained.
    \item Cross-polarization level of the whole system.
\end{itemize}

The simulations also allowed the determination of the level of beam spillover and clipping, thereby enabling an assessment of the noise contribution of the optical system. The procedure implemented follows that described in \cite{candotti10design}. Additionally, the contribution of the emission of mirrors due to their physical temperature ($T_{phys}$) was added to the source temperature noise as $T_{source,n} = T_{source}R_{so} + \varepsilon T_{phys} R_{so}$, where $T_{source}$ is the receiver noise temperature for the first mirror of the system. The receiver noise temperature for B6 is $T_{RX} = 50\,K$, as reported by \cite{ALMAhb2019}, which is practically the same temperature at the exit window of the LLAMA cryostat. Here, $R_{so}$ represents the relative power from the source falling on the reflector, and $\varepsilon$ denotes the emissivity of the mirror, with a value of 0.005 for the NACOS mirrors, and values of 0.003 and 0.02 for the sub-reflector and primary reflector, respectively, as stated by \cite{schwan2011invited}. The contribution to the noise temperature due to the emissivity of each mirror of NACOS is $\sim 1.5\,K$. With a total noise temperature contribution of $\sim 9\,K$ for receivers in Cab-A, and $\sim 11\,K$ for receivers in Cab-B ($\sim 18\,K$ for B1), NACOS will increase the observing time of LLAMA to achieve the same sensitivity as the receivers placed at the telescope focal plane, according to the radiometer equation  \cite{Kraus1986}, $\Delta t_{obs}=[(T_{RX}+T_{noise-NACOS}+T_{noise-Telescope})/(T_{RX}+T_{noise-Telescope})]^2$ (see Table \ref{table:Noise_temperature_contribution}).

\begin{figure}[H]
\begin{center}
\begin{tabular}{c}
\includegraphics[height=14cm]{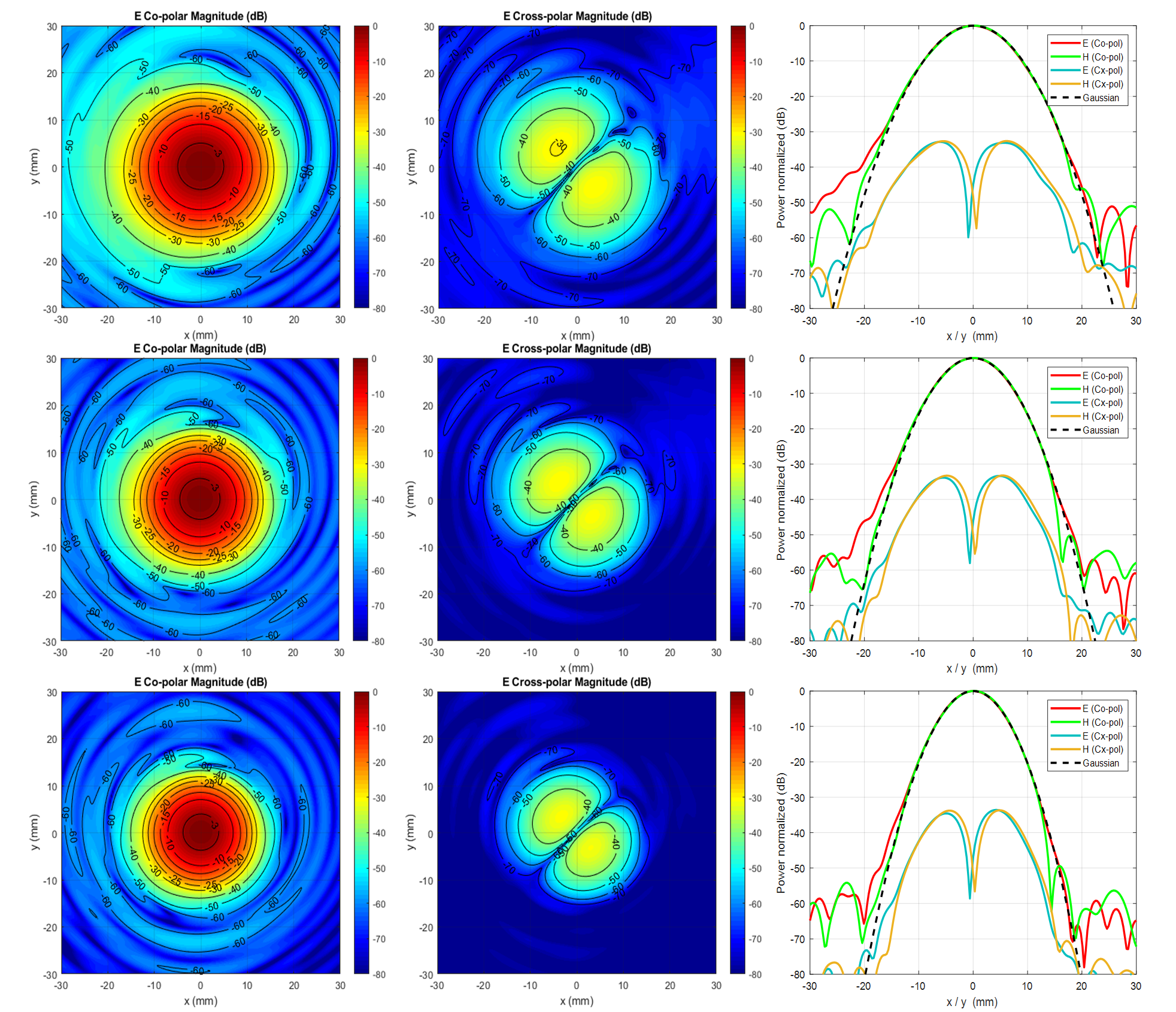}  
\\
(a) \hspace{4.8cm} (b) \hspace{4.8cm} (c)
\end{tabular}
\end{center}
\caption 
{ \label{fig:Cab_A_B6_Beam_pattern_NASS}
Plots of the beam pattern on the Nasmyth focal plane for B6 in Cab-A: (a) Co-polar plot. (b) Cross-polar plot. (c) Cut plot. Top row: 211\,GHz. Middle row: 243\,GHz. Bottom row: 275\,GHz.} 
\end{figure}

\begin{figure}[H]
	\centering
	\includegraphics[width=0.75\textwidth]{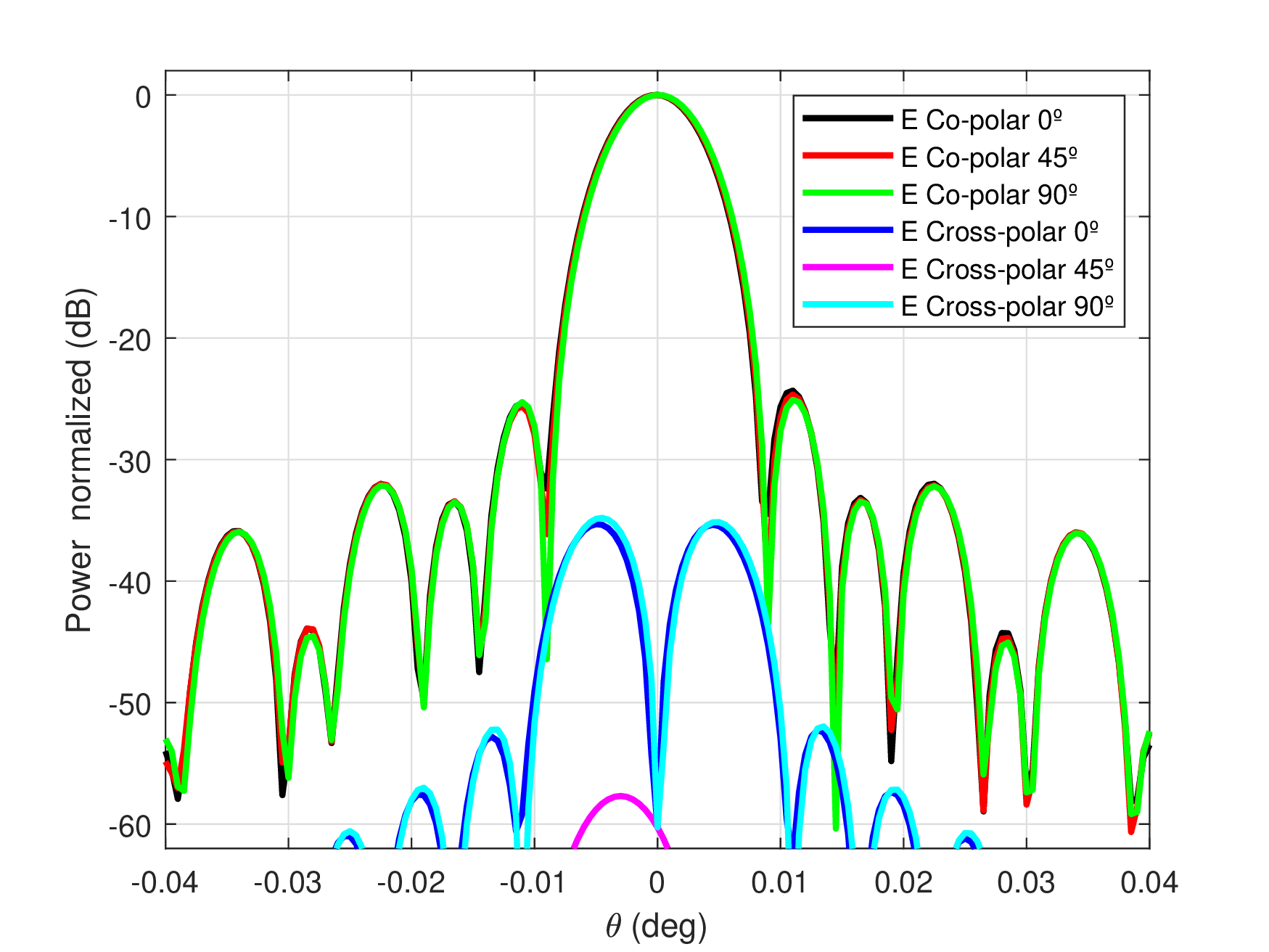}\caption{Antenna beam pattern on the sky for B6 (243\,GHz) in Cab-A.}
	\label{fig:Beam_patern_sky_CAB_A_3GBT_B6_MF}
\end{figure}

The PO analysis of B6 for Cab-A is presented as an example. The beam pattern on the Nasmyth focal plane (Fig. \ref{fig:Cab_A_B6_Beam_pattern_NASS}), the antenna beam pattern on the sky (Fig. \ref{fig:Beam_patern_sky_CAB_A_3GBT_B6_MF}), and the noise temperature contribution \ref{table:Noise_temperature_contribution_Cab_A_B6}  are presented. The results for all the bands of Cab-A and Cab-B, including both FL and LT configurations are summarized in Tables \ref{table:Optical_performance_of_the_tertiary_system}, \ref{table:Optical_performance_of_the_radiotelescope}, and \ref{table:Noise_temperature_contribution}.

\begin{table}[H]
\caption{Noise temperature contribution for B6 (243\,GHz) in Cab-A.\\
The $T_{phys}$ is 296\,K  and 300\,K inside and outside the antenna respectively and the Accum. power is the accumulated power throughout the optical path. The sub-reflector surrounding temperature is obtained from \cite{ASC}, with the telescope pointing to zenith. Temperatures with subscript $n$ are the noise contribution.} 
\centering
\scalebox{0.7}{\begin{tabular}{|ccccccccccc|}
\hline
& M3A-B6 & Flat 2-B6 &Flat 1-B6 &R Flat &M2A &Encoder (bf) & Encoder (ff) &M1A &Sub-ref. &Primary ref.  \\
$T_{source}$(K) &50 &51.505 &53.027 &54.568 &56.082 &57.601 & 57.683 & 57.743 &59.317 &64.422 \\
$T_{surr}$(K) &296 &296 &296 &296 &296 &296 & 296 & 296 &2.798* &300 \\
$T_{phys}$(K) &296 &296 &296 &296 &296 &296 &296 &296 &300 &300 \\
$\varepsilon$ &0.005 &0.005 &0.005 &0.005 &0.005 &0 &0 &0.005 &0.003 &0.02 \\
$R_{so}$ &0.99993 &0.99988 &0.99982 &0.99990 &0.99989 &0.99977 &0.99983 &0.99973 &0.93578 &0.98586 \\
Accum. power &0.99993 &0.99981 &0.99963 &0.99953 &0.99942 &0.99919 &0.99902 &0.99875 &0.93462 &0.92140 \\
$T_{source,n}$(K) &51.484 &52.991 &54.516 &56.053 &57.568 &57.615 & 57.693 & 59.238 &64.230 &71.262 \\
$T_{surr,n}$(K) &0.0216 &0.0361 &0.0518 &0.0284 &0.0335 &0.0684 & 0.0501 & 0.0794 &0.1920 &4.3037 \\
$T_{Refl,n-out}$(K) &51.505 &53.027 &54.568 &56.082 &57.601 &57.683 &57.743 & 59.317 &64.422 &75.565 \\
$T_{Refl,n-add}$(K) &1.505 &1.522 &1.541 &1.514 &1.520 &0.0820 &0.060 & 1.575 &5.105 &11.143 \\
Noise cont.(\%) &5.89 &5.95 &6.03 &5.92 &5.94 &0.32 &0.23 & 6.16 &19.97 &43.59 \\

\hline
\end{tabular}}
\label{table:Noise_temperature_contribution_Cab_A_B6}
\end{table}


\begin{table}[H]
\caption{Optical performance of the tertiary system.}
\centering
\scalebox{0.69}{\begin{tabular}{|cccccccccccc|} 
\hline
\multicolumn{12}{|c|}{\textbf{Gaussian beam launched from CASS to NASS}}                                                                                                                                                                                                                                                                                                                                                                                                                                                                                                                                                                                                                                                                                              \\
Band             & \begin{tabular}[c]{@{}c@{}}Freq.\\(GHz)\end{tabular} & Config.  & \begin{tabular}[c]{@{}c@{}}Beam \\coupling(\%)\end{tabular}  & \begin{tabular}[c]{@{}c@{}}Gaussicity\\(\%)\end{tabular} & \begin{tabular}[c]{@{}c@{}}$\omega_{0x}$\\(mm)\end{tabular} & \begin{tabular}[c]{@{}c@{}}$\omega_{0y}$\\(mm)\end{tabular} & \begin{tabular}[c]{@{}c@{}}$\omega_{0expect}$\\(mm)\end{tabular} & \begin{tabular}[c]{@{}c@{}}$x_{offset}$\\(mm)\end{tabular} & \begin{tabular}[c]{@{}c@{}}$y_{offset}$\\(mm)\end{tabular} & \begin{tabular}[c]{@{}c@{}}Ellipticity\\(\%)\end{tabular}& \begin{tabular}[c]{@{}c@{}}Cx-pol\\(dB)\end{tabular}  \\ 
\hline
\hline
&211                                                  &     & 99.74                                                                                                      & 99.75                                                    & 8.54                                                       & 8.55                                                       & 8.51                                                              & -0.0672                                                    & 0.0048                                                     & 0.16 & -29.48                                                   \\
B6 & 
243                                                  & Cab-A LT    & 99.80                                                                                                      & 99.81                                                    & 7.42                                                       & 7.42                                                       & 7.39                                                              & -0.0478                                                    & 0.0261                                                     & 0.01 & -30.20\\  
&275                                                  &     & 99.83                                                                                                      & 99.84                                                    & 6.56                                                       & 6.56                                                       & 6.53                                                              & -0.0339                                                    & 0.0405                                                     & 0.06 & -30.72\\
\hline
&275                                                  &     & 99.81                                                                                                      & 99.81                                                    & 6.57                                                       & 6.56                                                       & 6.53                                                              & -0.0257                                                    & 0.0000                                                      & 0.13 & -58.02                                                   \\
B7 &324                                                  & Cab-A LT     & 99.86                                                                                                      & 99.86                                                    & 5.57                                                       & 5.56                                                       & 5.54                                                              & -0.0304                                                    & 0.0000                                                      & 0.16 & -61.36\\  
&373                                                  &     & 99.88                                                                                                      & 99.88                                                    & 4.84                                                       & 4.83                                                       & 4.81                                                              & -0.0319                                                     & 0.0000                                                  & 0.18 & -62.15\\
\hline
&602                                                  &     & 99.88                                                                                                      & 99.88                                                    & 3.00                                                       & 3.00                                                       & 2.98                                                              & 0.0080                                                    & 0.0084                                                      & 0.10 & -32.25                                                   \\
B9 &661                                                  & Cab-A LT    & 99.88                                                                                                      & 99.89                                                    & 2.73                                                       & 2.73                                                       & 2.72                                                              & 0.0100                                                    & 0.0054                                                      & 0.08 & -32.36\\  
&720                                                  &     & 99.88                                                                                                      & 99.89                                                    & 2.51                                                       & 2.51                                                       & 2.49                                                              & 0.0115                                                    & 0.0031                                                      & 0.10 & -32.47\\
\hline
\hline
&35                                                  &  & 98.81                                                                                                      & 99.01                                                    & 48.86                                                       & 50.41                                                       & 50.76                                                              & 1.1284                                                    & 0.0780                                                      & 3.17 & -26.73                                                   \\
B1 &42.5                                                  & Cab-B LT       & 99.45                                                                                                      & 99.51                                                    & 42.55                                                       & 43.57                                                       & 42.54                                                              & 0.7730                                                    & 0.0209                                                      & 2.41 & -26.90\\  
&52                                                  &     & 99.78                                                                                                      & 99.78                                                    & 35.07                                                       & 35.48                                                       & 34.82                                                              & 0.0400                                                    & 0.0026                                                                                                       & 1.17 & -27.11\\                                                                                                                                                                                                      \hline
&67                                                  &     & 99.88                                                                                                      & 99.90                                                    & 27.87                                                       & 27.73                                                       & 27.39                                                              & 0.2416                                                    & 0.0000                                                     & 0.49 & -27.26                                                   \\
B2+3 &91                                                  & Cab-B LT    & 99.85                                                                                                      & 99.89                                                    & 20.97                                                       & 20.88                                                       & 20.70                                                              & 0.0366                                                    & 0.0000                                                                                                       & 0.46 & -27.41\\  
&116                                                  &     & 99.84                                                                                                      & 99.87                                                    & 16.30                                                       & 16.24                                                       & 16.12                                                              & 0.0233                                                    & 0.0000                                                      & 0.39 & -27.28\\
\hline
&163                                                  &     & 99.82                                                                                                      & 99.83                                                    & 11.17                                                       & 11.12                                                       & 11.02                                                              & 0.0252                                                    & 0.0145                                                      & 0.45 & -26.96                                                   \\
B5 &187                                                  & Cab-B FL/LT    & 99.81                                                                                                      & 99.82                                                    & 9.74                                                       & 9.71                                                       & 9.60                                                              & 0.0257                                                    & 0.0150                                                      & 0.31 & -26.96\\  
&211                                                  &     & 99.81                                                                                                      & 99.82                                                    & 8.63                                                       & 8.60                                                       & 8.51                                                              & 0.0260                                                    & 0.0153                                                      & 0.35 & -26.95\\
\hline
\hline
&211                                                  &     & 99.60                                                                                                      & 99.62                                                    & 8.66                                                       & 8.63                                                       & 8.51                                                              & 0.0548                                                    & 0.0000                                                      & 0.36 & -26.95                                                   \\
B6 &243                                                  & Cab-B FL    & 99.51                                                                                                      & 99.53                                                    & 7.53                                                       & 7.51                                                       & 7.39                                                              & 0.0543                                                    & 0.0000                                                      & 0.35 & -26.94\\  
&275                                                  &     & 99.41                                                                                                      & 99.44                                                    & 6.67                                                       & 6.64                                                       & 6.53                                                              & 0.0542                                                    & 0.0000                                                      & 0.35 & -26.94\\
\hline
&602                                                  &     & 98.07                                                                                                      & 98.10                                                    & 3.12                                                       & 3.12                                                       & 2.98                                                              & 0.0268                                                    & 0.0051                                                      & 0.19 & -26.93                                                   \\
B9 &661                                                  & Cab-B FL    & 97.71                                                                                                      & 97.74                                                    & 2.86                                                       & 2.86                                                       & 2.72                                                              & 0.0269                                                    & 0.0050                                                      & 0.17 & -26.96\\  
&720                                                  &     & 97.32                                                                                                      & 97.36                                                    & 2.65                                                       & 2.64                                                       & 2.49                                                              & 0.0271                                                    & 0.0049                                                      & 0.15 & -26.93\\
\hline
\end{tabular}}
\label{table:Optical_performance_of_the_tertiary_system}
\end{table}


\begin{table}[H]
\caption{Optical performance of the radiotelescope.}
\centering
\scalebox{0.71}{\begin{tabular}{|ccccccccc|} 
\hline
\multicolumn{9}{|c|}{\textbf{Gaussian beam launched from NASS to sky}}                                                                                                                                                                                                                                                                                                                                                                                                                                                                                                                                                                                                                                                                                              \\
Band &\begin{tabular}[c]{@{}c@{}}Freq.\\(GHz)\end{tabular} & Config.             & 
\begin{tabular}[c]{@{}c@{}}Ap. eff\\(\%)\end{tabular} &
\begin{tabular}[c]{@{}c@{}}Illum. over\\main refl.(dB)\end{tabular} & \begin{tabular}[c]{@{}c@{}}FWHM (arcsec)\\(Max ; Min)\end{tabular} &  \begin{tabular}[c]{@{}c@{}}Peak (arcsec)\\(Az ; El)\end{tabular} & \begin{tabular}[c]{@{}c@{}}Side lobes (dB)\\(left ; rigth)\end{tabular} & \begin{tabular}[c]{@{}c@{}}Cx-pol\\(dB)\end{tabular}   \\ 
\hline
\hline
&211                                                  
& 
& 78.42                                                           
& 10.77                                                       
& (28.76 ; 28.68)                                               
& (0.47 ; -0.04)                                                    
& (-25.10 ; -24.21)                                                       
& -31.67                                                                                                      \\
B6 &243                                                  
& Cab-A LT    
& 78.70                                                       
& 11.05                                                       
& (24.99 ; 24.95)                                               
& (0.40 ; 0.00)                                                    
& (-25.36 ; -24.30)                                                       
& -32.10
\\  
&275                                                  
&     
& 79.04                                                       
& 11.17                                                       
& (22.06 ; 22.03)                                               
& (0.32 ; 0.04)                                                    
& (-25.40 ; -24.24)                                                       
& -32.54
\\
\hline
&275                                                  
& 
& 79.01                                                           
& 11.15                                                       
& (22.06 ; 22.06)                                               
& (0.03 ; 0.00)                                                    
& (-24.38 ; -25.05)                                                       
& -66.70                                                                                                      \\
B7 &324                                                  
& Cab-A LT    
& 79.39                                                       
& 11.24                                                       
& (18.70 ; 18.69)                                               
& (0.03 ; 0.00)                                                    
& (-24.41 ; -24.92)                                                       
& -67.55
\\  
&373                                                  
&     
& 79.50                                                       
& 11.23                                                       
& (16.23 ; 16.22)                                               
& (0.04 ; 0.00)                                                    
& (-24.61 ; -24.98)                                                       
& -67.96
\\
\hline
&602                                                  
& 
& 79.69                                                           
& 11.27                                                       
& (10.04 ; 10.04)                                               
& (0.09 ; 0.04)                                                    
& (-24.67 ; -24.79)                                                       
& -34.06                                                                                                      \\
B9 &661                                                  
& Cab-A LT    
& 79.67                                                       
& 11.26                                                       
& (9.15 ; 9.14)                                               
& (0.08 ; 0.03)                                                    
& (-24.70 ; -24.88)                                                       
& -34.16
\\  
&720                                                  
&     
& 79.66                                                       
& 11.25                                                       
& (8.40 ; 8.40)                                               
& (0.07 ; 0.02)                                                    
& (-24.72 ; -24.93)                                                       
& -34.23
\\
\hline
\hline
&35                                                  
& 
& 67.88                                                           
& 6.26                                                       
& (171.00 ; 163.80)                                               
& (5.82 ; 0,00)                                                    
& (-20.32 ; -19.37)                                                       
& -30.97                                                                                                      \\
B1 &42,5                                                  
& Cab-B LT    
& 73.81                                                       
& 8.50                                                       
& (144.00 ; 138.00)                                               
& (3.80 ; 0.00)                                                    
& (-21.96 ; -20.35)                                                       
& -29.64
\\  
&52                                                  
&     
& 75.88                                                       
& 10.05                                                       
& (106.80 ; 115.20)                                               
& (1.78 ; 0,00)                                                    
& (-23.19 ; -21.51)                                                       
& -28.87
\\                                                                                          \hline
&67                                                  
& 
& 76.94                                                           
& 11.11                                                       
& (91.52 ; 91.39)                                               
& (0.18 ; 0.00)                                                    
& (-24.53 ; -26.28)                                                       
& -28.69                                                                                                      \\
B2+3 &91                                                  
& Cab-B LT    
& 77.29                                                       
& 11.93                                                       
& (67.89 ; 67.79)                                               
& (-0.12 ; 0.00)                                                    
& (-25.29 ; -26.20)                                                       
& -28.67
\\  
&116                                                  
&     
& 77.34                                                       
& 11.70                                                       
& (52.96 ; 52.93)                                               
& (0.10 ; 0.00)                                                    
& (-24.31 ; -25.68)                                                       
& -28.66
\\
\hline
&163                                                  
& 
& 78.89                                                           
& 11.19                                                       
& (37.22 ; 37.20)                                               
& (-0.14 ; -0.04)                                                    
& (-24.86 ; -24.87)                                                       
& -28.47                                                                                                      \\
B5 &187                                                  
& Cab-B FL/LT    
& 78.87                                                       
& 11.16                                                       
& (32.44 ; 32.43)                                               
& (-0.12 ; -0.04)                                                    
& (-24.89 ; -24.90)                                                       
& -28.48
\\  
&211                                                  
&     
& 78.85                                                       
& 11.16                                                       
& (28.76 ; 28.74)                                               
& (-0.11 ; -0.04)                                                    
& (-24.90 ; -24.92)                                                       
& -28.48
\\
\hline
\hline
&211                                                  
& 
& 78.87                                                           
& 11.20                                                       
& (28.77 ; 28.74)                                               
& (-0.18 ; 0.00)                                                    
& (-25.10 ; -24.78)                                                       
& -28.48                                                                                                      \\
B6 &243                                                  
& Cab-B FL    
& 78.84                                                       
& 11.19                                                       
& (24.98 ; 24.96)                                               
& (-0.16 ; 0.00)                                                    
& (-25.07 ; -24.71)                                                       
& -28.47
\\  
&275                                                  
&     
& 78.85                                                       
& 11.21                                                       
& (22.06 ; 22.06)                                               
& (-0.16 ; 0.00)                                                    
& (-25.05 ; -24.60)                                                       
& -28.47
\\
\hline
&602                                                  
& 
& 78.58                                                           
& 11.25                                                       
& (10.07 ; 10.06)                                               
& (-0.13 ; -0.01)                                                    
& (-25.06 ; -23.70)                                                       
& -28.45                                                                                                      \\
B9 &661                                                  
& Cab-B FL    
& 78.46                                                       
& 11.26                                                       
& (9.17 ; 9.17)                                               
& (-0.13 ; -0.01)                                                    
& (-24.47 ; -23.51)                                                       
& -28.45
\\  
&720                                                  
&     
& 78.31                                                       
& 11.27                                                       
& (8.42 ; 8.42)                                               
& (-0.13 ; -0.01)                                                    
& (-24.34 ; -23.31)                                                       
& -28.44
\\
\hline
\end{tabular}}
\label{table:Optical_performance_of_the_radiotelescope}
\end{table}

\begin{table}[H]
\caption{Noise temperature contribution for FL and LT configurations. $T_{noise-NACOS}$ is the noise temperature added by the NACOS system. $T_{noise-Telescope}$ is the noise temperature added by the primary and sub-reflector, considering an effective sky temperature obtained from \cite{ASC} with the telescope pointing to zenith. $\Delta t_{obs}$ is the increase in observing time attributed to NACOS.} 
\centering
\begin{tabular}{|ccccccc|}
\hline
\begin{tabular}[c]{@{}c@{}}Band\end{tabular} & \begin{tabular}[c]{@{}c@{}}Freq. \\(GHz)\end{tabular} & Config. & \begin{tabular}[c]{@{}c@{}}$T_{RX}$ \\(K)\end{tabular} & \begin{tabular}[c]{@{}c@{}}$T_{noise-NACOS}$ \\ (K)\end{tabular} & \begin{tabular}[c]{@{}c@{}}$T_{noise-Telescope}$\\ (K)\end{tabular}  & \begin{tabular}[c]{@{}c@{}}$\Delta t_{obs}$\\ (\%)\end{tabular} \\ 
\hline
B6  & 243  & Cab-A LT    & 50 & 9.317 & 16.248 & 30.1 \\  

B7  & 324  & Cab-A LT    & 72 & 9.102 & 21.893 & 20.3\\  

B9  & 661  & Cab-A LT    & 105 & 9.182 & 24.611 & 14.7\\  

B1  & 42.5  & Cab-B LT    & 28 & 17.637 & 18.503 & 90.2\\  

B2+3  & 91  & Cab-B LT    & 40 & 10.727 & 16.249 & 41.8\\  

B5  & 183  & Cab-B FL/LT    & 50 & 10.872 & 16.845 & 35.2\\  

B6  & 243  & Cab-B FL    & 50 & 10.778 & 16.149 & 35.2\\  

B9  & 661  & Cab-B FL    & 105 & 10.771 & 24.826 & 17.3\\  

\hline
\end{tabular}
\label{table:Noise_temperature_contribution}
\end{table}

\begin{figure}[H]
	\centering
	\includegraphics[width=0.95\textwidth]{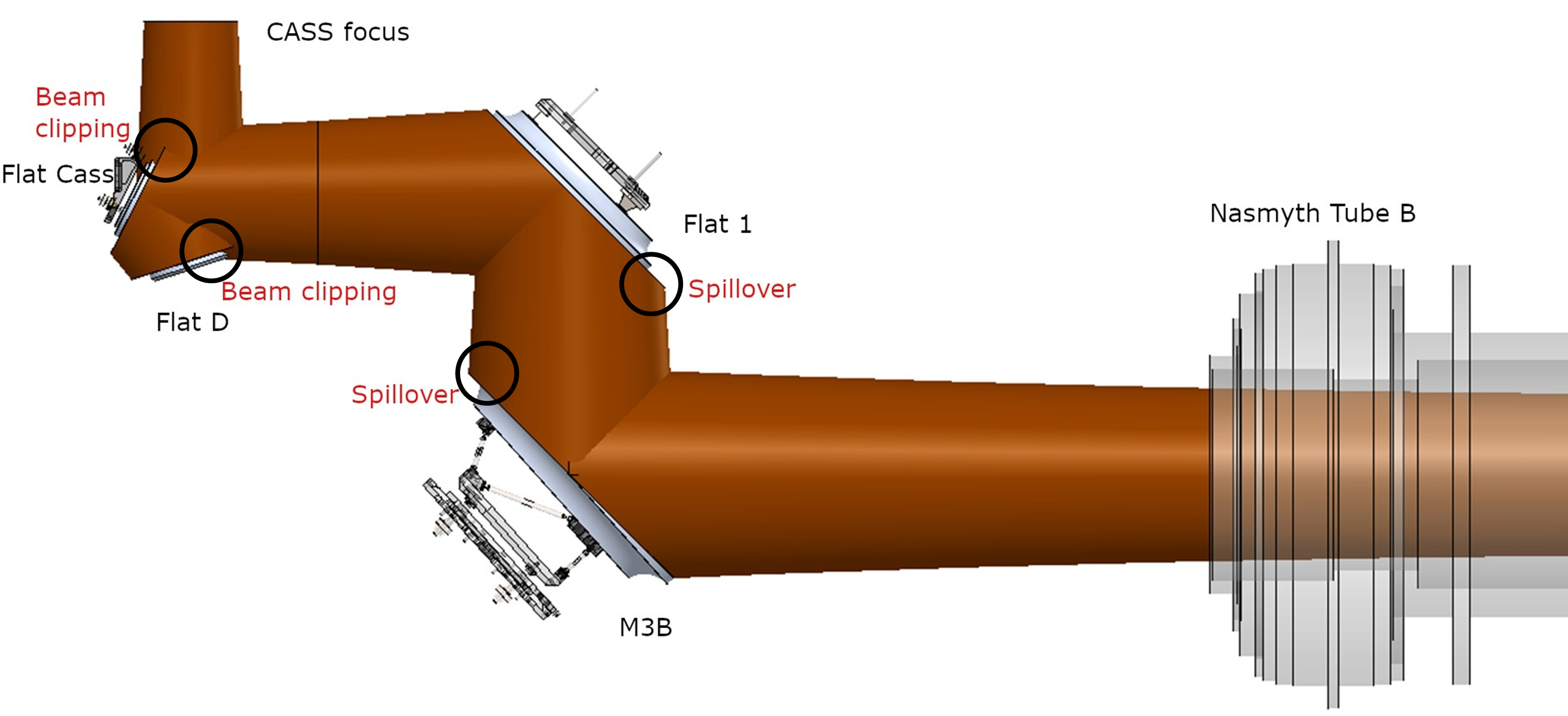}\caption{B1 beam clipping and spillover on the optical train of common mirrors.}
	\label{fig:B1_mirror_interference}
\end{figure}

Based on the data presented in Tables \ref{table:Optical_performance_of_the_tertiary_system}, \ref{table:Optical_performance_of_the_radiotelescope}, and \ref{table:Noise_temperature_contribution}, the following observations can be made:

\begin{itemize}
     \item For LT configuration, the beam coupling and Gaussicity are higher than 99.5\% for all cases except for B1. Losses and distortion for B1 are  attributed to beam clipping  while propagating through mirrors Flat D and Flat Cass, with an energy loss of $\sim 2\%$ at 35\,GHz and $\sim 1.2\%$ at 42.5\,GHz, and some distortion on the beam (ellipticity $\sim3\%$). The Nasmyth tube clips the 35\,GHz beam at 26\,dB (29\,dB at 42.5\,GHz). Additional losses of $\sim5\%$ at 35\,GHz and $\sim 2\%$ at 42.5\,GHz  is produced by beam spillover along the optical train of the common mirrors. $\sim 70\%$ of this spillover is produced at mirrors Flat D and Flat Cass (Fig. \ref{fig:B1_mirror_interference}). Attempts to reduce spillover in these two mirrors increase the beam clipping, keeping the energy losses almost constant.
       
    \item For FL, the beam coupling and Gaussicity of B9 is reduced due to the defocus loss mentioned at the end of Sec. \ref{NACOS-LT}.
    
    \item The aperture efficiency of B1 is reduced due to energy loss from beam spillover and interference along the common optics of the optical train..

    \item The contribution of NACOS to the system's noise temperature increases LLAMA's observing time by $\sim22\%$ for Cab-A and $\sim32\%$ for Cab-B ($\sim90\%$ for B1).
\end{itemize}

\section{Tolerance Analysis}
\label{Tolerance analysis}

Tolerance analysis in optical systems is essential to evaluate how misalignment of optical components and manufacturing tolerances can affect the overall performance of the optical system. Even in systems like LLAMA, where the optical components are adjustable in translations and tilts, this analysis is crucial to understand the impact of perturbations on optical characteristics. 

Misalignment of individual mirrors have different effects on the performance of the system. Flat mirrors with lateral offset have almost no effect on the optical performance, except for the amount of energy that spills over the offset mirror. On the other hand, curved mirrors with focal length $f$ will produced an angular deviation in the beam of $\sim \Delta /f$. An axial offset $\Delta$ in a tilted flat mirror will result in a $\Delta$ offset in the beam and an equal increment in its path length, leading to defocus. An axial offset $\Delta$ in a curved mirror will result in a defocus which also depends on $f$, according to the thin lens equation ($1/d_{in} + 1/d_{out} = 1/f$). A tilt $\alpha$ in any mirror results in a $2\alpha$ beam angular deviation. Misalignment of mirrors can be caused by both static factors (e.g., parts manufacture, assembly, mounting) and dynamic factors (e.g., telescope's EL position, temperature variations). In the present work, we focus only on static factors.

Degradation in aperture efficiency, errors in telescope pointing, and increases in cross-polarization levels are critical parameters for the telescope's performance. Additionally, increases in sidelobe levels and the FWHM size, though less critical, are also undesirable.

\begin{figure}[H]
	\centering
	\includegraphics[width=0.75\textwidth]{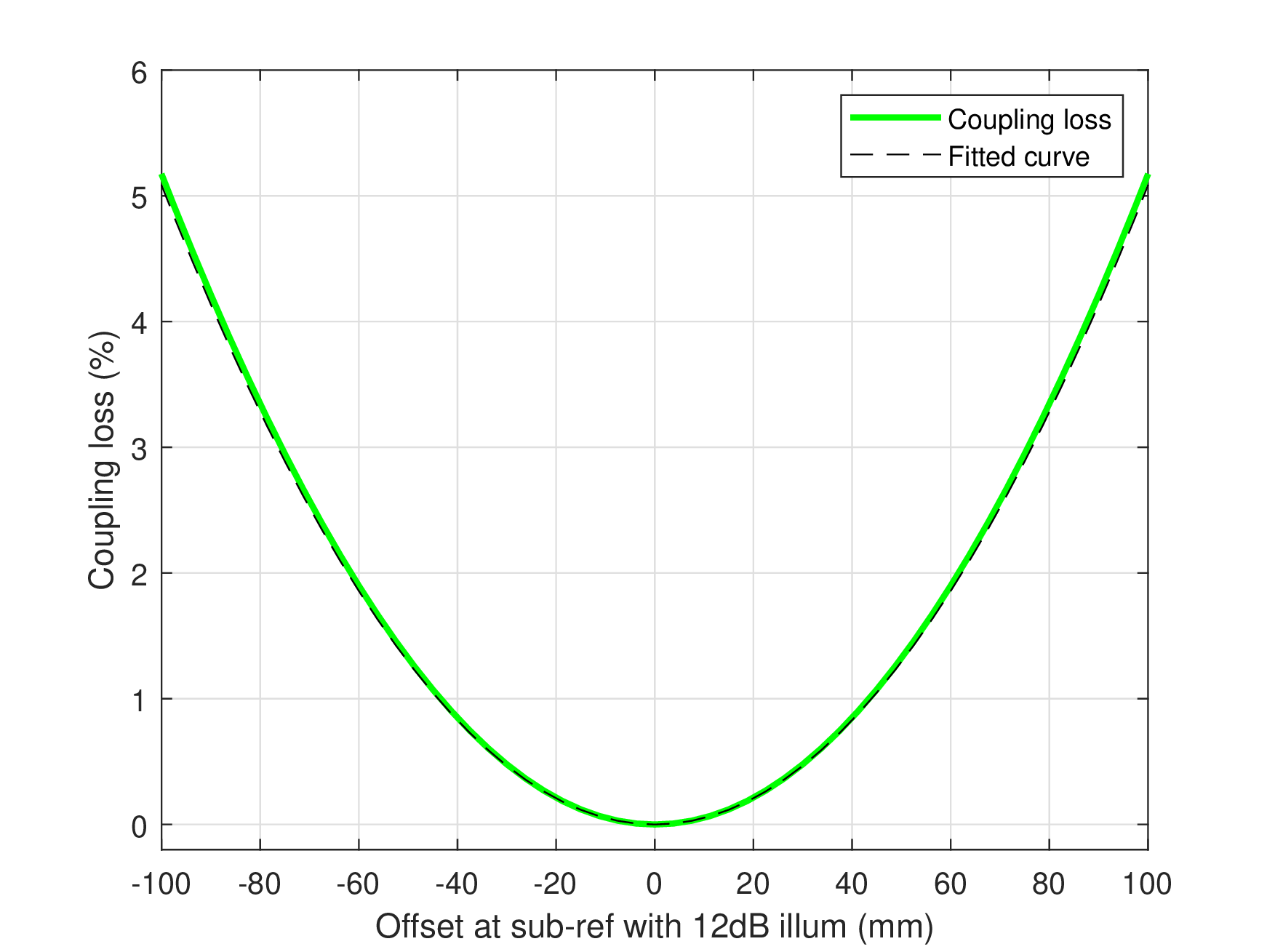}\caption{Coupling loss at the sub-reflector due to receiver's beam offset.}
	\label{fig:Coupling_loss_estimation}
\end{figure}

The decrease in aperture efficiency primarily occurs when the primary reflector is not symmetrically illuminated at its center, indicating a radial displacement in the aperture plane of the radio telescope \cite{holdaway2001alma}. This loss can be estimated using the coupling integral [Eq. (\ref{Eq:Gaussicity_orig})] onto the sub-reflector, considering the beam from the sky with an Airy pattern produced by the radio telescope aperture, and an offset fundamental Gaussian beam with a 12\,dB edge taper representing the receiver horn emission. The aperture efficiency loss estimation with respect to the beam offset (i.e., $x_{offset}$) at the sub-reflector is presented in Fig. \ref{fig:Coupling_loss_estimation} and accurately fitted with

\begin{equation}
Ap_{eff\:loss \: estim.} \left[\rm{\%}\right] = -1.17x10^{-9}(x_{offset}\left[\rm{mm}\right])^4 +0.00052(x_{offset}\left[\rm{mm}\right])^2.
\label{Eq:Ap_ef_loss_estimation}
\end{equation}

For LLAMA, the target in aperture efficiency loss is $\leq$ 1\%, although a loss of up to 2\% is considered acceptable.
According to Eq. (\ref{Eq:Ap_ef_loss_estimation}), $\sim$1\% loss is produced by an illumination offset at the sub-reflector of 41.3\,mm, equivalent to 11\% of its radius ($R_{s}$=375\,mm). This offset represents a tilt of the chief ray at the Cassegrain telescope focal plane of $0.4\,^{\circ}$ ($\arctan(41.3\,\rm mm / 5882.86\,\rm mm)$).

The precise pointing of the telescope beam is crucial for radio astronomical observations with high spatial resolution. While the telescope's movement control system primarily ensures pointing accuracy, any misalignment in the optical system can cause a deviation between the telescope's geometric and optical axes. In telescopes with a single receiver, this deviation can be corrected within the pointing algorithms. However, in telescopes like LLAMA, which incorporate multiple receivers sharing numerous optical components, especially with the ultimate goal of multi-band observation, the situation is more complex. Ensuring the correct alignment of the optical system is imperative to guarantee that the pointing in the sky is, within a certain margin, consistent for all receivers simultaneously. A pointing error of FWHM$/$10 results in a reduction of the intensity coupled by the telescope of $\sim3\%$, representing the accepted limit in most cases \cite{36FootTelescope, cwik1996beam, AlmaFeReq}.

For LLAMA, B9 is the band with highest resolution (FWHM=9.15\,arcsec at its central frequency of 661\,GHz), therefore, the maximum admissible pointing error is 0.915\,arcsec, which represents a lateral offset of the chief ray at the Cassegrain telescope focal plane of $0.43\,\rm mm$ ($\tan(0.915\,\rm arcsec) \cdot 96000\,\rm mm$).

Misalignment of the mirrors will also degrade the optical performance of the system due to defocusing and spillover, impacting the size and asymmetry of the telescope's main beam (i.e., FWHM), the intensity level of the side lobes, and the level of cross-polarization (Cx-pol).
The LLAMA requirements for these parameters are: FWHM size increment $\leq$2\%, sidelobe level $\leq$-22\,dB and Cx-pol efficiency $\geq$99.5\% (-23\,dB).

There are different analysis techniques to correlate mechanical uncertainties with deviations in the optical parameters of the system: The Worst Case Scenario (WCS), which consists of adding all the absolute maximums uncertainties to obtain in this way the worst level of expected optical performance; The Root Sum of Squares (RSS), which considers that the uncertainties of each variable are independent and can be described from a normal or Gaussian distribution ($\sigma_{system}=\sqrt{\sum_{i}^{n} \sigma^2_{i}}$). RSS provides results closer to the reality in comparison to WCS, but it does not provide information about the statistical behavior of the system. The Monte Carlo (MC) analysis \cite{prescott1965monte}, in which the uncertainties of each variable are characterized by a probability distribution, such that the variable has an associated probability of assuming a given value within a defined interval. The position and orientation of each component can then be modeled from a probability distribution, with a normal distribution used in this work. The mean value is set to the design value, and the standard deviation equals the manufacturing tolerance. By assigning a random value to each variable of the components based on its probability distribution, a real (probable) configuration of the system is generated, allowing the determination of optical performance. Conducting a large number ($\geq$ $10^3$) of individual calculations enables the statistical analysis of the system's optical performance based on the compilation of individual results.

In the current work, we performed a tolerance analysis of NACOS to evaluate how the fabrication tolerances of the mirrors impact the overall optical system performance. This was done by implementing the MC technique with two methodologies, each varying in complexity, computing time, and result accuracy:

\begin{enumerate}
    \item The ray tracing (RT) through the implementation of ABCD matrix formalism, in which the errors in position and orientation were incorporated into the matrices of individual components. This approach allowed for rapid and reasonably accurate calculations of optical parameters such as aperture efficiency and pointing.
    \item PO simulations in GRASP, where errors in position and orientation were incorporated into the coordinate system of each component. This method provided highly accurate results for aperture efficiency, pointing, cross-polarization, sidelobe levels, and FWHM.
\end{enumerate}

\subsection{Monte Carlo Implemented in RT}
\label{Monte Carlo implemented in RT}

The RT method using the ABCD matrices can be extended to scenarios where the system components exhibit small linear and/or angular deviations \cite{siegman1986lasers}. In a misaligned system confined to a plane, the pointing error ($\Delta r_{out}$ and $\Delta \theta_{out}$) of the beam after having propagated through a component, can be characterized by a misalignment matrix $M_{mis}$ multiplied by an error vector $\epsilon$ \cite{lazareff2001alignment}. 

\begin{equation}
\begin{bmatrix}
\Delta r_{out} \\
\Delta\theta_{out} \\
\end{bmatrix}
=M_{mis}.\epsilon
=\begin{bmatrix}
1-cos\left(2\theta\right) & 0 & sen\left(2\theta\right)  \\
-\frac{1}{f} & -2 & 0 \\
\end{bmatrix}
\left[\begin{matrix}
\Delta X \\
\Delta\alpha \\
\Delta Z \\
\end{matrix}\right].
\label{Eq: Misaligned_error_matrix}
\end{equation}

The $M_{mis}$ matrix incorporates the focal length of the component ($f$) and the angle of incidence concerning the nominal optical axis ($\theta$). The error vector $\epsilon$ encompasses both linear ($\Delta$X and $\Delta$Z) and angular ($\Delta\alpha$) alignment errors of the component.

\begin{figure}[H]
	\centering
	\includegraphics[width=0.8\textwidth]{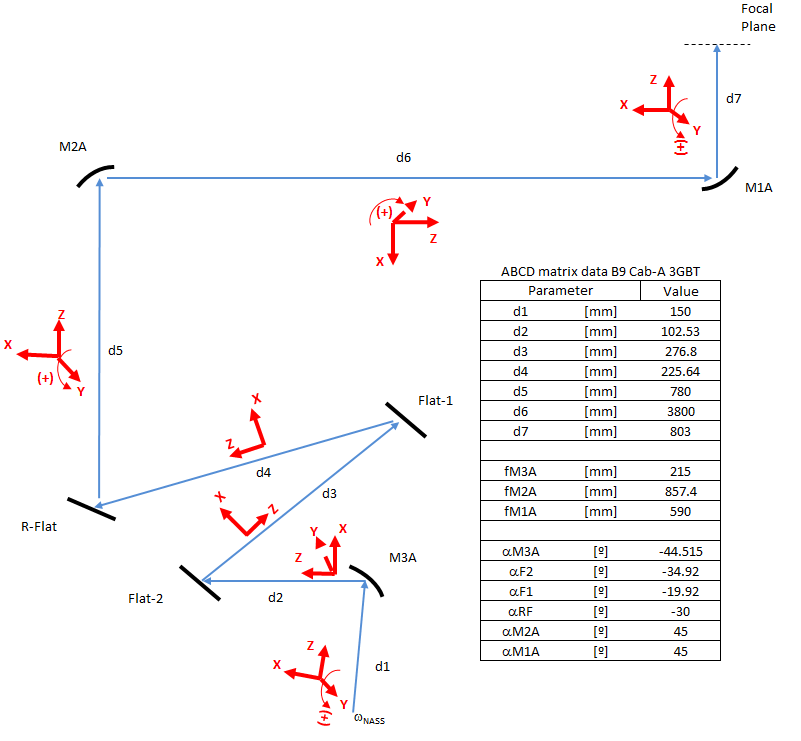}	
	\caption{RT schematic of B9 for Cab-A configuration.}
		\label{fig:B9_Cab-A_ABCD}
	\end{figure}

The error introduced by each optical component ($\epsilon_i$) in the system can be incorporated into the cascade multiplication of the ABCD matrix, as outlined in Sec. \ref{Tertiary optical system design}.  This approach enables tracking the path of an incoming beam ($B_{in}=[r_{in}; \theta_{in}]$) throughout the entire system, up to its exit ($B_{out}=[r_{out}; \theta_{out}]$).
As an example, the misaligned beam at the telescope focal plane ($B_{FP}=[r_{FP}; \theta_{FP}]$) resulting from the propagation of the input beam ($B_{\omega_{NASS}}=[r_{\omega_{NASS}} ; \theta_{\omega_{NASS}}]$ at the Nasmyth focal plane) along the B9 system for Cab-A configuration (Fig. \ref{fig:B9_Cab-A_ABCD}) is described by

\begin{equation}
\begin{split}
B_{FP} & = M_{B9} B_{\omega_{NASS}} + M_{d7} M_{M1A} M_{d6} M_{M2A} M_{d5} M_{RF} M_{d4} M_{F1} M_{d3} M_{F2} M_{d2} M_{M3A_{mis}} \epsilon_{M3A} \\
& + M_{d7} M_{M1A} M_{d6} M_{M2A} M_{d5} M_{RF} M_{d4} M_{F1} M_{d3} M_{F2_{mis}} \epsilon_{F2} \\ 
& + M_{d7} M_{M1A} M_{d6} M_{M2A} M_{d5} M_{RF} M_{d4} M_{F1_{mis}} \epsilon_{F1} \\ 
& + M_{d7} M_{M1A} M_{d6} M_{M2A} M_{d5} M_{RF_{mis}} \epsilon_{RF} \\ 
& + M_{d7} M_{M1A} M_{d6} M_{M2A_{mis}} \epsilon_{M2A} \\
& + M_{d7} M_{M1A_{mis}} \epsilon_{M1A},
\end{split}
\label{Eq:B9_misaligned_ABCD_matrix}
\end{equation}
where $M_{B9}$ is the matrix that describes the aligned optical system and is given by

\begin{equation}
M_{B9} = M_{d7} M_{M1A} M_{d6} M_{M2A} M_{d5} M_{RF} M_{d4} M_{F1} M_{d3} M_{F2} M_{d2} M_{M3A} M_{d1}, \nonumber\\
\label{Eq:B9_aligned_ABCD_matrix}
\end{equation}
being $M_j$ and $M_{di}$ the matrices correspondent to each j-mirror and i-distance respectively [Eq. (\ref{Ec: Matrice})]. 

The offset and tilt values obtained at the focal plane of the telescope (i.e., $B_{FP}$) can then be projected onto the sub-reflector plane to estimate the aperture efficiency loss using Eq. (\ref{Eq:Ap_ef_loss_estimation}).

A script was written in which every component of Eq. (\ref{Eq:B9_misaligned_ABCD_matrix}) takes random values for $\Delta$X, $\Delta$Z and $\Delta\alpha$ in each run. These values were obtained from a normal distribution with a mean of zero and $6\sigma$ equivalent to the mechanical tolerance range.
For LLAMA, each mirror has three assembly interfaces: the mirror to its support, the support to the mirror assembly support, and the assembly support to the NACOS structure  (CASS or NASS). The manufacturing assembly tolerance for each instance is $\pm$0.1\,mm ($\pm$0.05\,º) for linear (angular) displacements. Hence, the tolerance of each mirror assembly is $tol_{assy}=\pm\sqrt{3tol^2_{i}}$ ($\equiv \pm3\sigma$ which guarantees that 99.73\% of the values are inside tolerance). Thus, $\sigma_X$ = $\sigma_Z$ = 0.058\,mm and $\sigma_{\alpha}$ = 0.029º. 

\begin{figure}[H]
\begin{center}
\begin{tabular}{c}
\includegraphics[height=15.5cm]{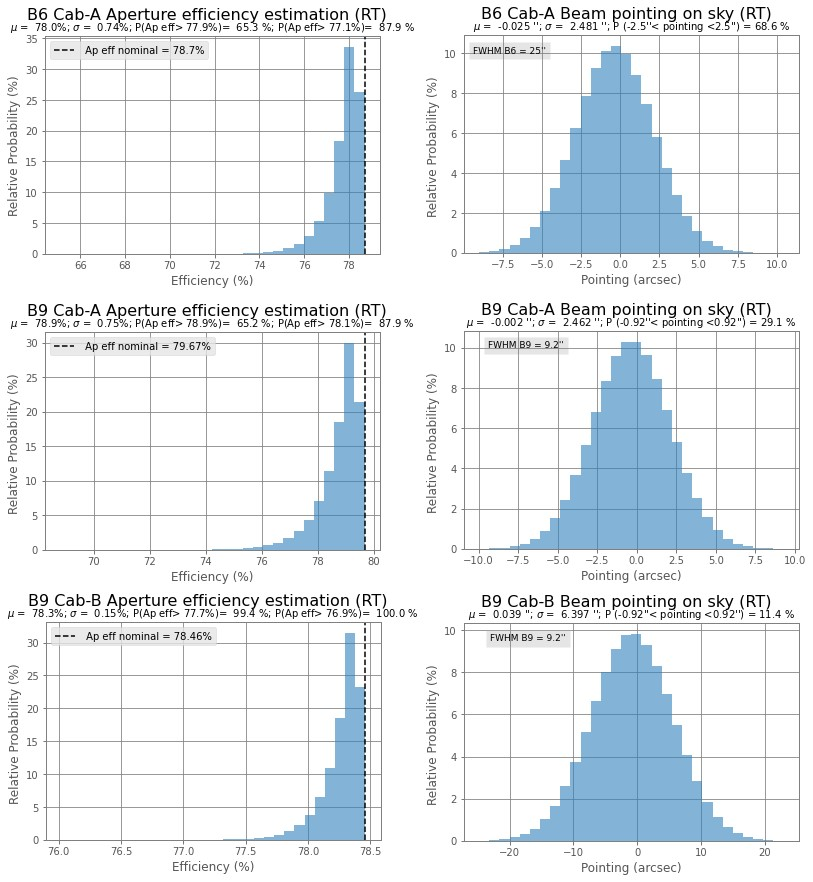}  
\\
(a) \hspace{7.1cm} (b)
\end{tabular}
\end{center}
\caption 
{ \label{fig:MC_RT_plots}
Distributions of: (a) Aperture efficiency estimation and (b) pointing. Top row: B6 for Cab-A. Middle row: B9 for Cab-A. Bottom row: B9 for Cab-B.} 
\end{figure}

A total of 10$^4$ runs were performed for B6 and B9 for Cab-A LT configuration, and B9 for Cab-B FL configuration. The aperture efficiency and pointing distributions are shown in Fig. \ref{fig:MC_RT_plots}. Each plot includes the nominal value and the probability of meeting the LLAMA requirements, derived from the best-fit distribution determined using the \textit{Generalized Akaike Information Criterion}\cite{rigby2019distributions}.

In conclusion, the Monte Carlo analysis applied to RT shows that tolerances have a minor impact on aperture efficiency. There is a 35\% and 12\% probability of degrading efficiency by more than 1\% and 2\% respectively, for both B6 and B9 in Cab-A. The degradation is negligible for B9 in Cab-B.
Regarding the pointing, B9 for Cab-A has a probability of only 29\% of satisfying the acceptance criterion of being within FWHM$/$10. Although B6 presents the same dispersion ($\sigma\sim$2.5\,arcsec) in the pointing as B9, since FWHM of B6 is greater, the probability of satisfying the acceptance criterion is higher (69\%). The pointing of B9 for Cab-B has a probability of 11\% of satisfying the acceptance criterion.

\subsection{Monte Carlo Implemented in GRASP}
\label{Monte Carlo implemented in GRASP}

\begin{figure}[H]
	\centering
	\includegraphics[width=0.65\textwidth]{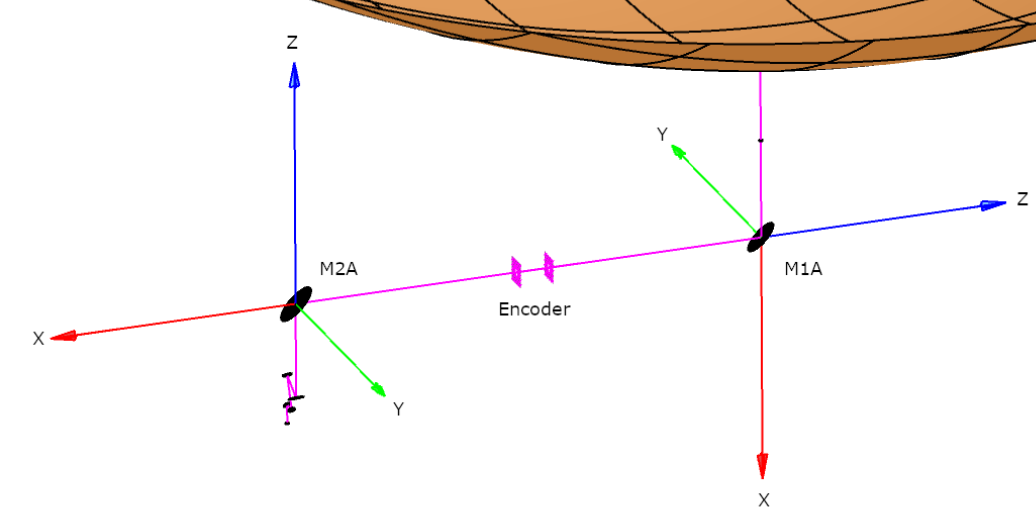}	
	\caption{Coordinate system used for mirror misalignment on GRASP (M1A and M2A shown as example).}
		\label{fig:Coordinates_for_Monte_Carlo_PO}
	\end{figure}

The Monte Carlo method was applied through 10$^3$ simulations using PO in GRASP for B6 and B9 of Cab-A for LT configuration. Each simulation introduced random misalignments consistent with those described in Sec. \ref{Monte Carlo implemented in RT}). Misalignment in position ($\Delta$X, $\Delta$Y and $\Delta$Z) and orientation rotations around the X, Y, Z axes ($\Delta$a, $\Delta$b and $\Delta$c) were applied to the local coordinate systems of each mirror, as illustrated in Fig. \ref{fig:Coordinates_for_Monte_Carlo_PO}.

\begin{figure}[H]
\begin{center}
\begin{tabular}{c}
\includegraphics[height=15.5cm]{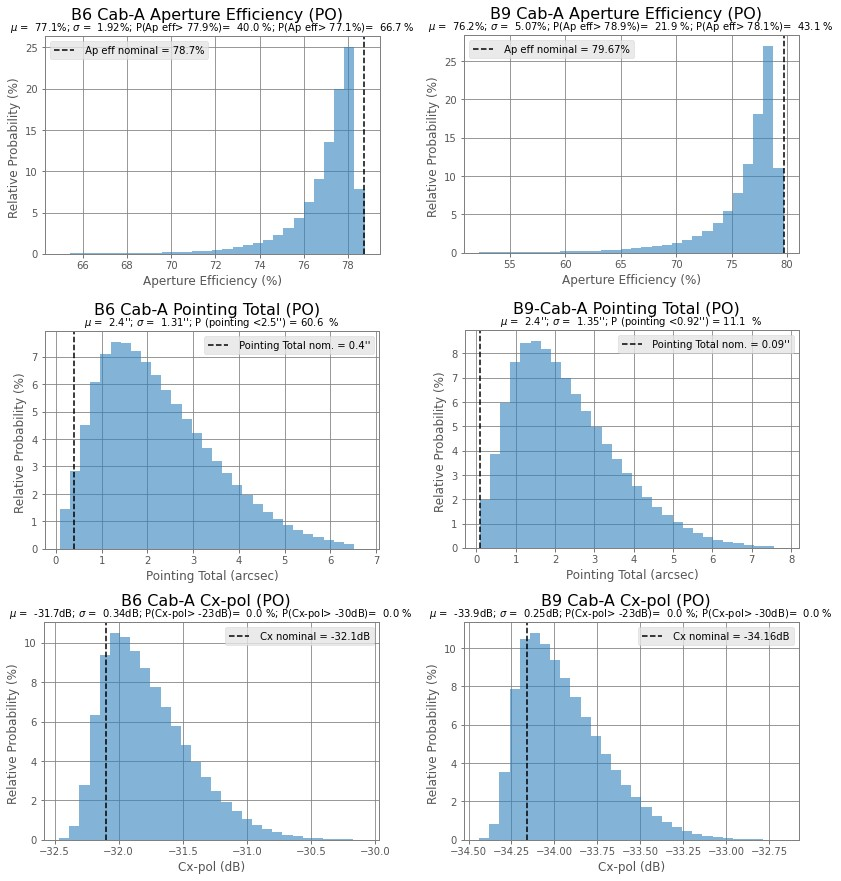}  
\\
(a) \hspace{7.1cm} (b)
\end{tabular}
\end{center}
\caption 
{ \label{fig:MC_PO_plots_Ap_poin_CX}
Distributions of: (a) B6 for Cab-A (b) B9 for Cab-A. Top row: Aperture efficiency. Middle row: Pointing. Bottom row: Cx-pol level.} 
\end{figure}

\begin{figure}[H]
\begin{center}
\begin{tabular}{c}
\includegraphics[height=21.6cm]{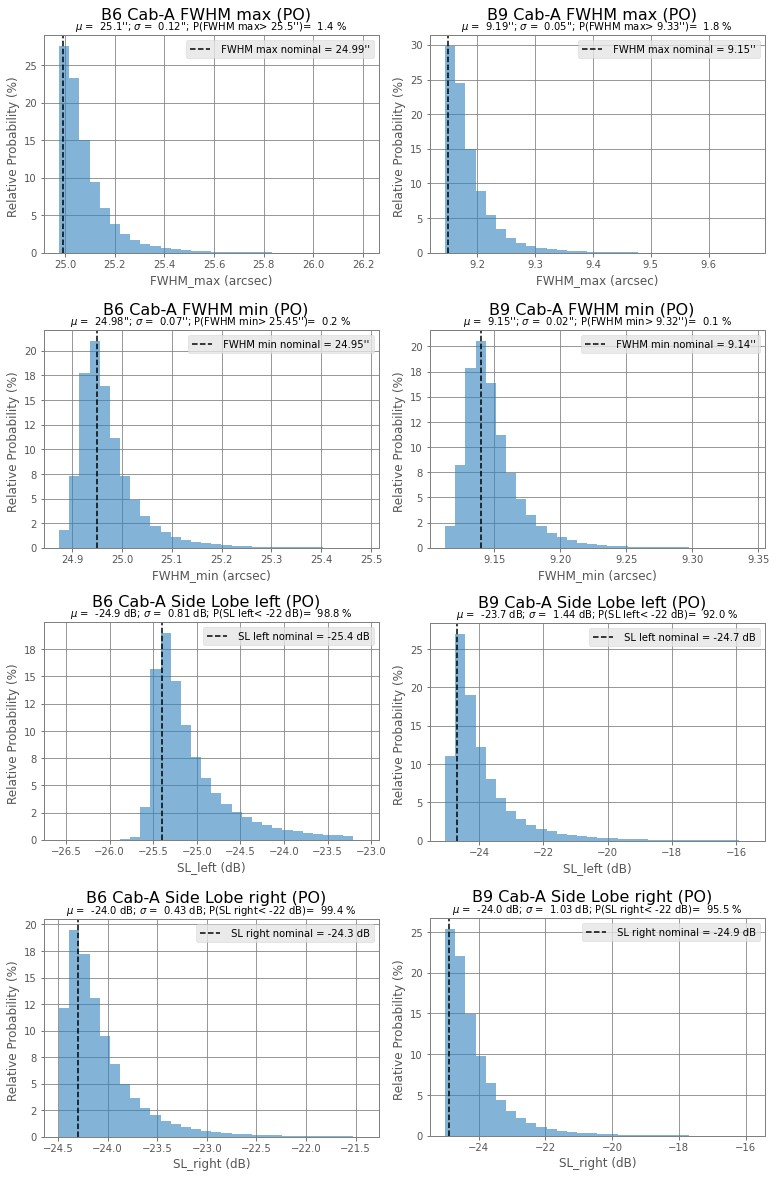}  
\\
(a) \hspace{7.1cm} (b)
\end{tabular}
\end{center}
\caption 
{ \label{fig:MC_PO_plots_FWHM_SL.png}
Distributions of FWHM and first side lobe level of: (a) B6 for Cab-A (b) B9 for Cab-A.} 
\end{figure}

\begin{figure}[H]
	\centering
	\includegraphics[width=1\textwidth]{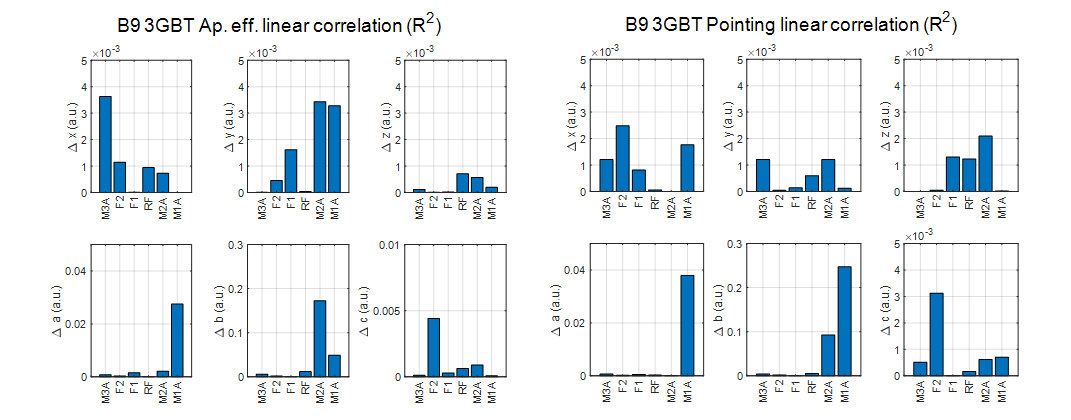}	
	\caption{Linear correlation between aperture efficiency and pointing with respect to mirror misalignment for B9 in Cab-A.}
		\label{fig:B9_linear_and_angular_correlation}
	\end{figure}

Aperture efficiency, pointing and Cx-pol distributions are shown in Fig. \ref{fig:MC_PO_plots_Ap_poin_CX}. Distributions for FWHM (maximum and minimum values) and first side lobe level (left and right with respect to the main lobe) are displayed in Fig. \ref{fig:MC_PO_plots_FWHM_SL.png}. The probability of aperture efficiency degradation exceeding 1\% is 60\% for B6 and $\sim$78\% for B9. The likelihood of degradation beyond 2\% is $\sim$23\% for B6 and $\sim$67\% for B9. The pointing shows a probability of fulfilling the acceptance criterion of $\sim$61\% for B6 and $\sim$11\% for B9. The effect of the misalignment of the mirrors in the Cx-pol, FWHM and first sidelobe level performance is negligible. 
Hence, B9 is the most sensitive channel to misalignment of the optical system. In Fig. \ref{fig:B9_linear_and_angular_correlation}, the linear correlation between aperture efficiency and pointing for B9 (the most sensitive channel of the system) with the linear and angular deviations of each mirror is presented. Tilt with respect to Y-axis of both M1A and M2A mirrors presents the highest correlation. 

In conclusion, while misalignment has a negligible effect on FWHM, sidelobe levels, and Cx-pol, the MC analysis using PO confirms the Ray Tracing RT prediction of a very low probability, particularly for B9, of meeting the pointing requirement. Additionally, degradation in aperture efficiency is expected to exceed the requirement. Therefore, a precise alignment strategy was developed to improve the overall optical performance of NACOS-LT.

\begin{figure}[H]
\begin{center}
\begin{tabular}{c}
\includegraphics[height=7.0cm]{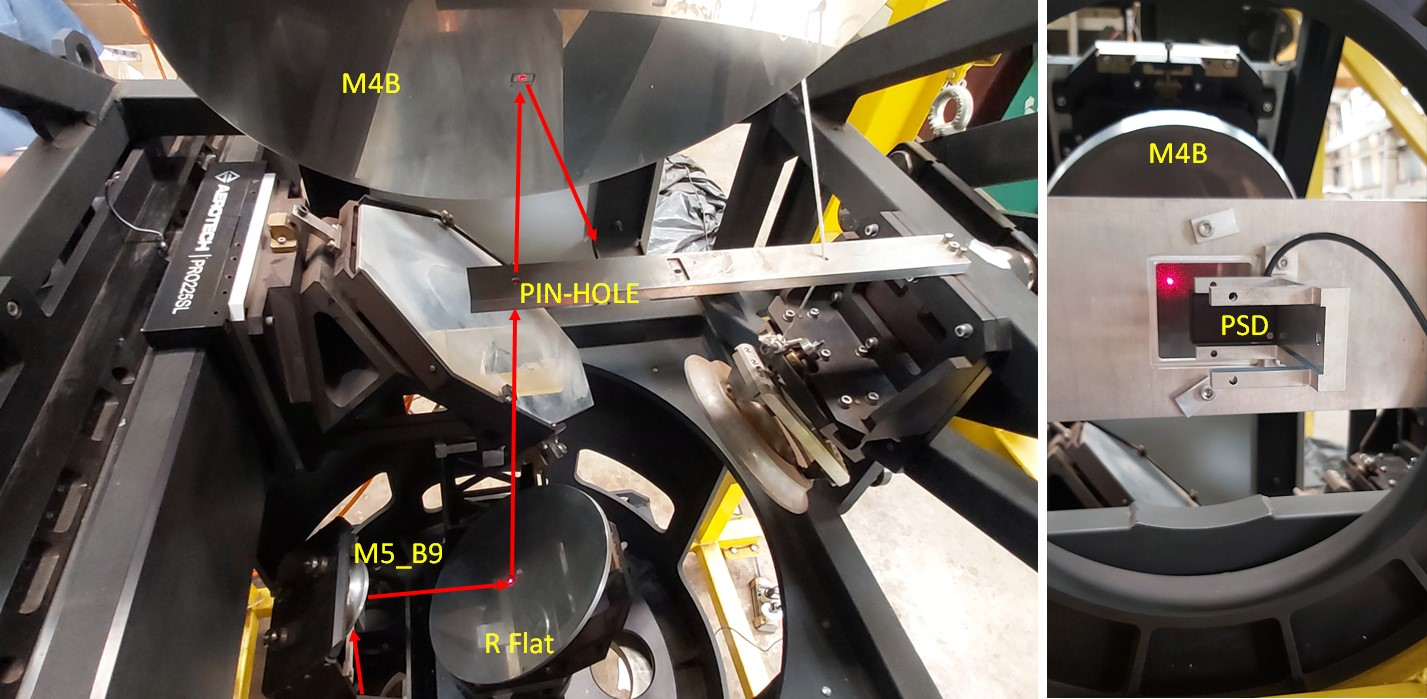}  
\\
\hspace{3.0cm} (a) \hspace{7.1cm} (b)
\end{tabular}
\end{center}
\caption 
{ \label{fig:Alignment_FL}
(a) Laser beam propagation along NASS during B9 alignment  (b) PSD measurement of the output laser beam from NASS.} 
\end{figure}

This alignment strategy consists of two stages. First, a coarse alignment is performed by propagating a laser beam through each subsystem (CASS or NASS) and visually evaluating it at a measurement plane. The mirrors are adjusted to ensure the laser beam passes through pinholes placed along the optical path, as shown in Fig. \ref{fig:Alignment_FL} (a). NACOS powered mirrors can be manually adjusted in X, Y, Z, $\Delta$a, and $\Delta$b, whereas the flat ones in Z, $\Delta$a, and $\Delta$b. 

The second stage involves fine-tune alignment using a positioning sensor device (PSD), placed at the measurement plane as shown in Fig. \ref{fig:Alignment_FL} (b). This device tracks the laser beam with an accuracy of $\sim$0.01\,mm.

This alignment strategy was implemented during the AIV stage of NACOS-FL. B9 achieved a pointing accuracy of 0.68\,arcsec and an estimated aperture efficiency loss of 0.1\%.

The proposed alignment method will be implemented again during the integration of NACOS in the radio telescope and will also serve as tool for realigning NACOS whenever necessary.  

Although NACOS may experience some dynamic misalignment due to gravitational effects from changes in the pointing of the radio telescope and/or temperature variations in the cabins, these are expected to have only a minor impact on system performance.
The optical components expected to be affected by gravitational effects are those located in Cab-CASS, as this cabin is the only one that moves with changes in the elevation angle of the radio telescope. The Nasmyth cabins are unaffected by this movement.

Structural simulations of the support assembly and mirror M1A showed maximum deformations of 0.08\,mm in position and 0.01\,deg in orientation. Evaluating this misalignment in the RT model results in a pointing error of $\sim$0.7\,arcsec ($\sim$8\% of FWHM), and an estimated aperture efficiency loss of 0.2\% for B9. Simulations predict a similar response for the optical components of NACOS for Cab-B.

Regarding temperature variations, the cabins housing CASS and NASS are equipped with a thermal control system maintaining temperatures within ±1\,K. Therefore, negligible deformations are expected.

\section{Conclusions}
\label{Conclusions}

A complete quasi-optical design has been developed for the long-term phase of the tertiary optical system (NACOS) of the LLAMA radio telescope. This was achieved by meeting the requirement to reuse most of the components of the optical system implemented for the first light phase of the project. Physical optics simulations were employed to complete the validation and characterization of the optical performance of the system.
Despite a minimal frequency dependence exhibited by the two lower-frequency receivers (i.e., B1 and B2+3), a frequency independent system has been achieved for the remaining receivers (i.e., B5, B6, B7, B9) of the radio telescope. The antenna aperture efficiency obtained by considering a fundamental Gaussian beam illumination was greater than 77\% for all the receivers, except for B1, which reached $\sim$68\% at its lower frequency because of a truncation in the beam due to the reuse of mirrors from the first light optical system. These values are not taking into account surface errors on reflectors. The design ensures aberration-free beams for all receivers, with Gaussicity levels over 99\%.

Additionally, a tolerance analysis was carried out to evaluate the degradation in the optical performance of the system caused by errors in the positioning of the optical components during the NACOS integration process. The analysis revealed that implementing an alignment strategy for the system is essential. Such a strategy was developed and successfully implemented during the assembly, integration, and verification phase of NACOS-FL.

\appendix    

\section{Determination of B1 and B2+3 focii Position and Size}
\label{Apendix1}

\begin{figure}[H]
	\centering
	\includegraphics[width=0.28\textwidth]{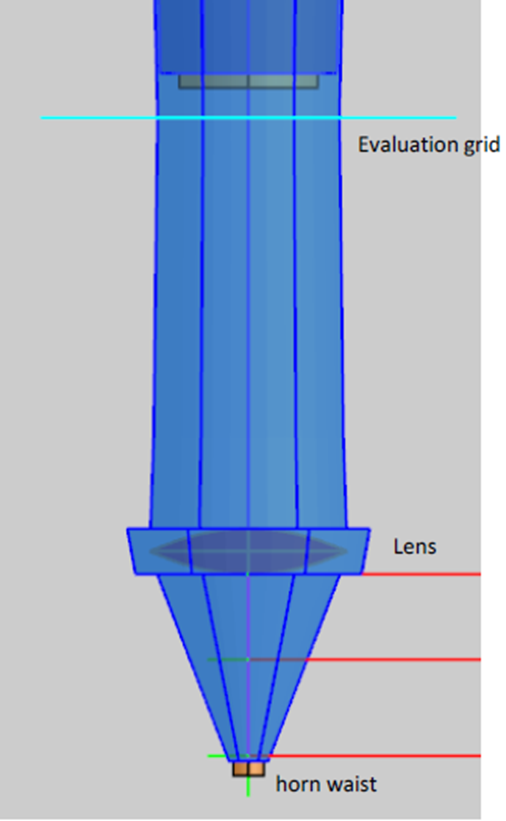}	
	\caption{GRASP set up for the evaluation of the output beam waist of the horn-lens system.}
		\label{fig:B1_evaluation_grid}
	\end{figure}

The size and position of the output beam waist from the horn-lens system were determined by simulating the propagation of the fundamental Gaussian beam from the phase location center (PLC) inside the horn (Fig. \ref{fig:B1_evaluation_grid}). This beam was traced through the lens, and the resulting electromagnetic field was evaluated in planes perpendicular to the optical axis at various positions. This process identified the location where the beam reached its smallest size ($\omega_{out} \equiv \omega_{NASS}$). The resulting beam waist parameters were then used as input for designing the optical system for the B1 and B2+3 band.

Table \ref{Summary_of_parameters_and_output_beam_characteristics} summarized the parameters used for the simulations and the position and size obtained of the output beam waist for the extreme and central frequencies of each band respectively.

\begin{table}[H]
\caption{B1 and B2+3 horn-lens optical parameters and output beam characteristics.}
\centering
\scalebox{0.9}{\begin{tabular}{|c|ccc|ccc|}
\hline
\multicolumn{1}{|c|}  {} & \multicolumn{3}{c|}  {B1} & \multicolumn{3}{c|}  {B2+3}   \\ \hline
Parameter & 35\,GHz  & 42.5\,GHz    & 52\,GHz    & 67\,GHz  & 91\,GHz  & 116\,GHz  \\ \hline
\multicolumn{1}{|r|}{PLC (mm)}   & 1.39 & 5.21 & 7.45  & 1.50 & 5.50 & 3.80 \\
\multicolumn{1}{|r|}{$\omega_{horn}$ (mm)}   & 9.40 & 9.38 & 9.40  & 4.80 & 4.80 & 4.80 \\
\multicolumn{1}{|r|}{lens material}   & \multicolumn{3}{c|}{HDPE} & \multicolumn{3}{c|}{HDPE} \\
\multicolumn{1}{|r|}{Refraction index n}   & \multicolumn{3}{c|}{1.5259} & \multicolumn{3}{c|}{1.5259} \\
\multicolumn{1}{|r|}{Loss tangent $\delta$}   & \multicolumn{3}{c|}{$2.73^{-4}$} & \multicolumn{3}{c|}{$2.73^{-4}$} \\
\multicolumn{1}{|r|}{Focal length f (mm)}   & \multicolumn{3}{c|}{181} & \multicolumn{3}{c|}{91} \\
\multicolumn{1}{|r|}{Lens diameter (mm)}   & \multicolumn{3}{c|}{190} & \multicolumn{3}{c|}{92} \\
\multicolumn{1}{|r|}{Lens thickness (mm)}   & \multicolumn{3}{c|}{40.8} & \multicolumn{3}{c|}{21.38} \\
\multicolumn{1}{|r|}{Lens shape}   & \multicolumn{3}{c|}{bi-hyperbolic} & \multicolumn{3}{c|}{bi-hyperbolic} \\
\multicolumn{1}{|r|}{distance lens to horn aperture (mm)}   & \multicolumn{3}{c|}{175.4} & \multicolumn{3}{c|}{81} \\
\multicolumn{1}{|r|}{$\omega_{NASS}$ (mm)}   & 50.76 & 42.54 & 34.82 & 27.39 & 20.70 & 16.12 \\
\multicolumn{1}{|r|}{Distance wrt cryo top plate (mm)}   & 563 & 553 & 473 & 90.50 & 131.80 & 80.50 \\
\multicolumn{1}{|r|}{$Te$ over the primary reflector (dB)}   & 11.74 & 12.16 & 12.20 & 12.53 & 13.20 & 13.01 \\
\hline
\end{tabular}}
\label{Summary_of_parameters_and_output_beam_characteristics}
\end{table}

\section{QO parameters of B6 for Cab-A propagation}
\label{Apendix2}

The QO parameters obtained from the propagation of B6 for Cab-A configuration are presented in Table \ref{table:CAB-A B6 Parámetros ópticos diseño 3GBT}.

\begin{figure}[H]
\begin{center}
\begin{tabular}{c}
\includegraphics[height=5.8cm]{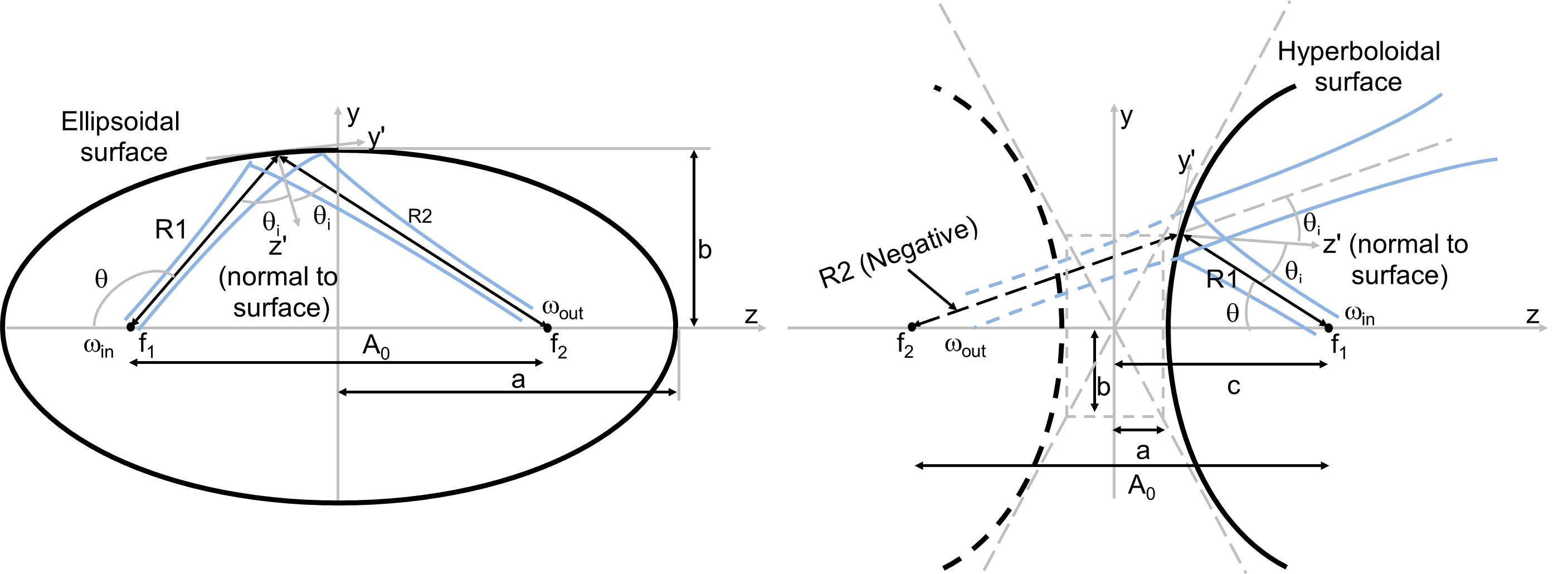}  
\\
(a) \hspace{7.0cm} (b)
\end{tabular}
\end{center}
\caption 
{ \label{fig:Conic sections}
Conic sections: (a) Ellipsoid, (b) Hyperboloid.} 
\end{figure}

The parameters that define the conical surface of each focusing mirror are shown in Fig. \ref{fig:Conic sections} and detailed in Table \ref{table:Mirror_geometrical_parameters_3GBT}. For the common mirrors M1A and M2A, the chosen parameters correspond to the central frequency (i.e., 452\,GHz) of the entire broadband range (i.e., 211 to 720\,GHz), as this minimizes loss due to phase distortion \cite{withington1992design} in the system.


\begin{table}[H]
\caption{QO parameters of B6 for CAB-A configuration.} 
\centering
\scalebox{0.75}{\begin{tabular}{|rccccc|}
\hline
\multicolumn{6}{|c|}{Fundamental GBP CAB-A B6}\\
\hline
\hline
Freq. (GHz) & 211 &227 &243 &259 &275 \\
$\omega_{NASS}\, \textrm{(mm)}$ &8.58	&7.97	&7.45	&6.99	&6.58\\
$d_{NASS-TOP PLATE}\, \textrm{(mm)}$ &-14.85	&-14.94	&-15.01	&-15.07	&-15.13\\
$d_{TOP PLATE-M3AB6}\, \textrm{(mm)}$ & \multicolumn{5}{c|}{\textsl{135}}\\
\hline
$f_{M3AB6}\, \textrm{(mm)}$ & \multicolumn{5}{c|}{\textsl{215}}\\
$\theta_{M3AB6}\,\hspace{0.5cm}
 \textrm{(deg)}$ & \multicolumn{5}{c|}{\textsl{43.81}}\\
$R_{IN M3AB6}\, \textrm{(mm)}$	&326.27	&302.27	&282.88	&266.99	&253.79\\
$\omega_{M3AB6}\, \textrm{(mm)}$	&11.67	&11.23	&10.87	&10.56	&10.30\\
$R_{OUT M3AB6}\, \textrm{(mm)}$	&-630.44	&-744.66	&-859.96	&-1104.17	&-1406.55\\
$K_{F\, M3AB6}\, \hspace{1cm}$	&0.99966	&0.99969	&0.99971	&0.99972	&0.99974\\
$K_{CO\, M3AB6}\, \hspace{1cm}$	&0.99932	&0.99937	&0.99941	&0.99944	&0.99947\\
\hline
$\omega_{0 OUT M3AB6}\, \textrm{(mm)}$	&10.53	&10.42	&10.30	&10.19	&10.06\\
$d_{OUT-M3AB6}\, \textrm{(mm)}$ &116.84	&103.92	&90.64	&77.08	&63.33\\
$d_{M3AB6-M2A}\, \textrm{(mm)}$ & \multicolumn{5}{c|}{\textsl{1385}}\\	
\hline
$f_{M2A}\, \textrm{(mm)}$ & \multicolumn{5}{c|}{\textsl{857.4}}\\
$\theta_{M2A}\,\hspace{0.5cm}
 \textrm{(deg)}$ & \multicolumn{5}{c|}{\textsl{45}}\\	
$R_{IN M2A}\, \textrm{(mm)}$ &1315.48	&1333.06	&1350.76	&1368.45	&1386.03\\
$\omega_{M2A}\, \textrm{(mm)}$	&55.52	&52.76	&50.43	&48.43	&46.71\\
$R_{OUT M2A}\, \textrm{(mm)}$ &-2462.21	&-2402.91	&-2347.47	&-2295.88	&-2248.03\\
$K_{F\, M2A}\, \hspace{1cm}$	&0.99948	&0.99953	&0.99957	&0.99960	&0.99963\\
$K_{CO\, M2A}\, \hspace{1cm}$	&0.99895	&0.99905	&0.99914	&0.99920	&0.99926\\
\hline
$\omega_{0 OUT M2A}\, \textrm{(mm)}$ &18.88	&18.01	&17.20	&16.44	&15.74\\
$d_{OUT-M2A}\, \textrm{(mm)}$ &2177.56	&2123.05	&2074.49	&2031.27	&1992.75\\
$d_{M2A-M1A}\, \textrm{(mm)}$ & \multicolumn{5}{c|}{\textsl{3800}}\\	
\hline
$f_{M1A}\, \textrm{(mm)}$ & \multicolumn{5}{c|}{\textsl{590}}\\
$\theta_{M1A}\,\hspace{0.5cm}
 \textrm{(deg)}$ & \multicolumn{5}{c|}{\textsl{45}}\\
$R_{IN M1A}\, \textrm{(mm)}$ &2004.48	&2031.26	&2053.69	&2072.63	&2088.73\\
$\omega_{M1A}\, \textrm{(mm)}$ &43.24	&43.12	&43.02	&42.94	&42.87\\
$R_{OUT-M1A}\, \textrm{(mm)}$ &-836.10	&-831.52	&-827.83	&-824.79	&-822.26\\
$K_{F\, M1A}\, \hspace{1cm}$	&0.99933	&0.99933	&0.99934	&0.99934	&0.99934\\
$K_{CO\, M1A}\, \hspace{1cm}$	&0.99866	&0.99866	&0.99867	&0.99868	&0.99868\\
\hline
$\omega_{CASS}\, \textrm{(mm)}$ &8.58	&7.97	&7.45	&6.99	&6.58\\
$d_{OUT-CASS}\, \textrm{(mm)}$ &803.19	&803.09	&803.01	&802.94	&802.88\\
$\omega_{CASS\,IDEAL}\, \textrm{(mm)}$ &8.51	&7.91	&7.39	&6.93	&6.53\\
$K_{AXIALL}\, \hspace{1cm}$	&0.99994	&0.99994	&0.99994	&0.99994	&0.99994\\
\hline
$d_{M1A-SUB REF}\, \textrm{(mm)}$ & \multicolumn{5}{c|}{\textsl{6685.86}}\\
$R_{IN SUB REF}\, \textrm{(mm)}$ &5887.16	&5886.65	&5886.24	&5885.90	&5885.62\\
$\omega_{SUB REF VERTEX}\, \textrm{(mm)}$ &310.50	&310.49	&310.48	&310.47	&310.47\\
$\omega_{SUB REF RIM}\, \textrm{(mm)}$ &316.37	&316.36	&316.35	&316.34	&316.33 \\
\hline
$K_{TOTAL}\, \hspace{1cm}$	&\textbf{0.99535}& \textbf{0.99558}& \textbf{0.99577}&	\textbf{0.99593}&	\textbf{0.98606}\\
$T_{e \hspace{0.15cm} SUB REF RIM}\, \textrm{(dB)}$ &\textbf{12.20}& \textbf{12.21}& \textbf{12.21}&	\textbf{12.21}&	\textbf{12.21}\\
$\eta_{APERTURE}\, \textrm{(\%)}$ &\textbf{79.89}& \textbf{79.89}& \textbf{79.89}&	\textbf{79.89}&	\textbf{79.89}\\
\hline
\end{tabular}}
\label{table:CAB-A B6 Parámetros ópticos diseño 3GBT}
\end{table}

\begin{table}[H]
\caption{Geometrical parameters of mirrors for LT in Cab-A configuration. Ell.: Ellipsoid , Hyp.: Hyperboloid.}
\centering
\scalebox{0.72}{\begin{tabular}{|c|cccccccccccc|}
\hline
\multicolumn{13}{|c|}{\textbf{MIRROR M1A}}                                                                                                                                                                                                                                                                                                                                                                                                                                   \\
\hline

\begin{tabular}[c]{@{}c@{}}Freq.\\ (mm)\end{tabular} &
\begin{tabular}[c]{@{}c@{}}$R_{in}$ \\(mm)\end{tabular} &
\begin{tabular}[c]{@{}c@{}}$R_{out}$ \\(mm)\end{tabular} & 
\begin{tabular}[c]{@{}c@{}}$f$ \\ (mm)\end{tabular} & 
\begin{tabular}[c]{@{}c@{}}$\theta_i$\\ (deg)\end{tabular}  &
\begin{tabular}[c]{@{}c@{}}$a$ \\ (mm)\end{tabular} &
\begin{tabular}[c]{@{}c@{}}$A_0$ \\ (mm)\end{tabular} &
\begin{tabular}[c]{@{}c@{}}$e$ \end{tabular} &
\begin{tabular}[c]{@{}c@{}}$b$ \\ (mm)\end{tabular} &
\begin{tabular}[c]{@{}c@{}}$\theta$ \\ (deg)\end{tabular} &
\begin{tabular}[c]{@{}c@{}}$\omega_m$ \\ (mm)\end{tabular} &
\begin{tabular}[c]{@{}c@{}}Diam. \\ (mm)\end{tabular} & 
\begin{tabular}[c]{@{}c@{}}Mirror \\  \end{tabular} \\ 
\hline

452                          & 809.95                    & 2172.62                   & 590.00                   & 45                            & 1491.29                  & 2318.69                   & 0.777411            & 938.01                   & 110.45  & 42.88                                                                                &277x178 & Ell.                                    \\
\hline

\multicolumn{13}{|c|}{\textbf{MIRROR M2A}}                                                                                                                                                                                                                                                                                                                                                                                                                                   \\
\hline

\begin{tabular}[c]{@{}c@{}}Freq.\\ (mm)\end{tabular} &
\begin{tabular}[c]{@{}c@{}}$R_{in}$ \\(mm)\end{tabular} &
\begin{tabular}[c]{@{}c@{}}$R_{out}$ \\(mm)\end{tabular} & 
\begin{tabular}[c]{@{}c@{}}$f$ \\ (mm)\end{tabular} & 
\begin{tabular}[c]{@{}c@{}}$\theta_i$\\ (deg)\end{tabular}  &
\begin{tabular}[c]{@{}c@{}}$a$ \\ (mm)\end{tabular} &
\begin{tabular}[c]{@{}c@{}}$A_0$ \\ (mm)\end{tabular} &
\begin{tabular}[c]{@{}c@{}}$e$ \end{tabular} &
\begin{tabular}[c]{@{}c@{}}$b$ \\ (mm)\end{tabular} &
\begin{tabular}[c]{@{}c@{}}$\theta$ \\ (deg)\end{tabular} &
\begin{tabular}[c]{@{}c@{}}$\omega_m$ \\ (mm)\end{tabular} &
\begin{tabular}[c]{@{}c@{}}Diam. \\ (mm)\end{tabular} & 
\begin{tabular}[c]{@{}c@{}}Mirror \end{tabular} \\ 
\hline

452                          & 1903.26                   & 1560.30                   & 857.40                   & 45                            & 1731.78                  & 2461.08                   & 0.710565            & 1218.54                  & 140.66                                                                                   &37.18  &329x223 & Ell.    \\                                 
\hline

\multicolumn{13}{|c|}{\textbf{MIRROR M3A}}                                                                                                                                                                                                                                                                                                                                                                                                                                   \\
\hline

\begin{tabular}[c]{@{}c@{}}Freq.\\ (mm)\end{tabular} &
\begin{tabular}[c]{@{}c@{}}$R_{in}$ \\(mm)\end{tabular} &
\begin{tabular}[c]{@{}c@{}}$R_{out}$ \\(mm)\end{tabular} & 
\begin{tabular}[c]{@{}c@{}}$f$ \\ (mm)\end{tabular} & 
\begin{tabular}[c]{@{}c@{}}$\theta_i$\\ (deg)\end{tabular}  &
\begin{tabular}[c]{@{}c@{}}$a$ \\ (mm)\end{tabular} &
\begin{tabular}[c]{@{}c@{}}$A_0$ \\ (mm)\end{tabular} &
\begin{tabular}[c]{@{}c@{}}$e$ \end{tabular} &
\begin{tabular}[c]{@{}c@{}}$b$ \\ (mm)\end{tabular} &
\begin{tabular}[c]{@{}c@{}}$\theta$ \\ (deg)\end{tabular} &
\begin{tabular}[c]{@{}c@{}}$\omega_m$ \\ (mm)\end{tabular} &
\begin{tabular}[c]{@{}c@{}}Diam. \\ (mm)\end{tabular} & 
\begin{tabular}[c]{@{}c@{}}Mirror \end{tabular} \\ 
\hline

243                          & 939.62                    & 278.79                    & 215.00                  & 43.81                         & 609.21                   & 968.95                    & 0.795253            & 369.35                   & 163.29                                          & 10.87        &68x47                                  & Ell.                                     \\
324                          & 6426.44                   & 222.44                    & 215.00                  & 44.515                        & 3324.44                  & 6426.52                   & 0.966557            & 852.56                   & 178.02                                          & 9.71         &60x42                                  & Ell.                                              \\
661                          & -755.76                   & 167.38                    & 215.00                  & 44.515                        & 294.19                   & 776.84                    & 1.320299            & 253.62                   & 76.59                                           & 8.42         &50x35                                  & Hyp.                                     \\
\hline
\end{tabular}}

\label{table:Mirror_geometrical_parameters_3GBT}
\end{table}

\subsection* {Code and Data Availability}
The data utilized in this study were obtained by LLAMA Collaboration. Data are available
from the authors upon request and with permission from LLAMA Collaboration.

\subsection* {Acknowledgments}
Emiliano Rasztocky extends his heartfelt gratitude to the staff of the Instituto Argentino de Radioastronomía for their invaluable support, which was fundamental in facilitating the publication of this work.\\
Rodrigo Reeves acknowledges support from ANID CATA BASAL FB210003.


\bibliography{report}   

\begin{thebibliography}{10}

\bibitem{arnal2017llama}
E.~Arnal, Z.~Abraham, C.~Cappa, {\em et~al.}, ``Llama: A new mm and submm observing facility,'' in {\em Revista Mexicana de Astronomia y Astrofisica Conference Series},   {\bf 49}, 53--53  (2017).

\bibitem{llamaweb}
{LLAMA web page}, ``\texttt{https://www.llamaobservatory.org/en/},''  (10 de {A}ugust de 2023).

\bibitem{romero2020large}
G.~E. Romero, ``Large latin american millimeter array,'' {\em arXiv preprint arXiv:2010.00738}   (2020).
\newblock [doi: https://doi.org/10.52712/sciencereviews.v1i4.10].

\bibitem{lepine2021llama}
J.~R. Lepine, Z.~Abraham, C.~G.~G. CASTRO, {\em et~al.}, ``The llama brazilian-argentinian radiotelescope project: progress in brazil and brics collaboration,'' {\em Anais da Academia Brasileira de Ci{\^e}ncias} {\bf 93}  (2021).
\newblock [doi: 10.1590/0001-3765202120200846].

\bibitem{fernandez2023llama}
M.~Fernandez-Lopez, P.~Benaglia, S.~Cichowolski, {\em et~al.}, ``Llama millimeter and submillimeter observatory. update on its science opportunities,'' {\em arXiv preprint arXiv:2312.12210}   (2023).
\newblock [doi:10.48550/arXiv.2312.12210].

\bibitem{gusten2006atacama}
R.~G{\"u}sten, L.~Nyman, P.~Schilke, {\em et~al.}, ``The atacama pathfinder experiment (apex)--a new submillimeter facility for southern skies--,'' {\em Astronomy \& Astrophysics} {\bf 454}(2), L13--L16  (2006).
\newblock [doi:10.1051/0004-6361:20065420].

\bibitem{johnson2023key}
M.~D. Johnson, K.~Akiyama, L.~Blackburn, {\em et~al.}, ``Key science goals for the next-generation event horizon telescope,'' {\em Galaxies} {\bf 11}(3), 61  (2023).
\newblock [doi:10.48550/arXiv.2304.11188].

\bibitem{almaweb}
{ALMA web page}, ``\texttt{https://www.almaobservatory.org/en/home/},''  (10 de {A}ugust de 2023).

\bibitem{LLAMAmanual}
V.~A. GmbH, ``{Large Latin-American Millimetric Array Telescope LLAMA – Antenna with Nasmyth Focus – Technical proposal.},'' tech. rep., Vertex Antennentechnik GmbH, VA Proj-No.: 21/09087  (Duisburg, 2014).

\bibitem{wootten2000frequency}
A.~Wootten, L.~Snyder, E.~van Dishoeck, {\em et~al.}, ``Frequency band considerations and recommendations,'' {\em NRAO, ALMA Millimeter Array Memo Series} (213)  (2000).

\bibitem{bachiller2000report}
R.~Bachiller, G.~Blake, R.~Booth, {\em et~al.}, ``Report of the alma scientific advisory committee: September 2000 meeting,''  (2000).

\bibitem{carter2007alma}
M.~Carter {\em et~al.}, ``Alma front-end optics design report,'' {\em available from ALMA project documentation server, Tech. Rep. FEND-40.02. 00.00-035-B-REP}   (2007).

\bibitem{carter2004alma}
M.~C. Carter, A.~Baryshev, M.~Harman, {\em et~al.}, ``Alma front-end optics,'' in {\em Ground-based Telescopes},   {\bf 5489}, 1074--1084, SPIE  (2004).
\newblock [doi:10.1117/12.552538].

\bibitem{lamb2001alma}
J.~Lamb, A.~Baryshev, M.~Carter, {\em et~al.}, ``Alma memo 362,''  (2001).

\bibitem{lamb2003low}
J.~W. Lamb, ``Low-noise, high-efficiency optics design for alma receivers,'' {\em IEEE Transactions on antennas and propagation} {\bf 51}(8), 2035--2047  (2003).
\newblock [doi:10.1109/TAP.2003.814743].

\bibitem{paul1998goldsmith}
F.~Paul, ``Goldsmith quasioptical systems: Gaussian beam quasioptical propogation and applications,''  (1998).

\bibitem{nystrom2009optics}
O.~Nystr{\"o}m, I.~Lapkin, V.~Desmaris, {\em et~al.}, ``Optics design and verification for the apex swedish heterodyne facility instrument (shefi),'' {\em Journal of Infrared, Millimeter, and Terahertz Waves} {\bf 30}, 746--761  (2009).
\newblock [doi:10.1007/s10762-009-9493-7].

\bibitem{gonzalez2014optics}
A.~Gonzalez, T.~Soma, T.~Shiino, {\em et~al.}, ``Optics characterization of a 900-ghz heb receiver for the aste telescope: design, measurement and tolerance analysis,'' {\em Journal of Infrared, Millimeter, and Terahertz Waves} {\bf 35}, 743--758  (2014).
\newblock [doi:10.1007/s10762-014-0074-z].

\bibitem{rioja2010precise}
M.~Rioja and R.~Dodson, ``Precise radio astrometry and new developments for the next-generation of instruments. aapr 2020, 28, 6,'' {\em arXiv preprint astro-ph.IM/2010.02156} .
\newblock [doi:10.1109/ICEAA.2019.8879329].

\bibitem{8440767}
M.~Kotiranta, K.~Jacob, H.~Kim, {\em et~al.}, ``Optical design and analysis of the submillimeter-wave instrument on juice,'' {\em IEEE Transactions on Terahertz Science and Technology} {\bf 8}(6), 588--595  (2018).

\bibitem{gonzalez2015design}
A.~Gonzalez, ``Design of optics for alma receivers at naoj,'' in {\em 36th ESA Antenna Workshop, Noordwijk, the Netherlands},   (2015).

\bibitem{murphy1987distortion}
J.~Murphy, ``Distortion of a simple gaussian beam on reflection from off-axis ellipsoidal mirrors,'' {\em International journal of infrared and millimeter waves} {\bf 8}, 1165--1187  (1987).
\newblock [doi:10.1007/BF01010819].

\bibitem{pontoppidan2008electromagnetic}
K.~Pontoppidan, ``Electromagnetic properties and optical analysis of the alma antennas and front ends,'' {\em Nat. Radio Astron. Observatory, Charlottesville, VA, ALMA EDM Doc. FEND-80.04. 00.00-026-A-REP}   (2008).

\bibitem{Tapia2015}
V.~Tapia, {\em {Design and measurements of an optical system for ALMA band 1}}.
\newblock PhD thesis  (2015).

\bibitem{Finger}
R.~Finger responsable del desarrollo~de B2+3, Universidad de~Chile, ``Comunicaciones personales,''  (2022).

\bibitem{gonzalez2016frequency}
A.~Gonzalez, ``Frequency independent design of quasi-optical systems,'' {\em Journal of Infrared, Millimeter, and Terahertz Waves} {\bf 37}(2), 147--159  (2016).
\newblock [doi:10.1007/s10762-015-0205-1].

\bibitem{ticraweb}
{TICRA web page}, ``\texttt{https://www.ticra.com/},''  (10 de {O}ctober de 2023).

\bibitem{johansson1995comparison}
J.~F. Johansson, ``A comparison of some feed types,'' in {\em Multi-Feed Systems for Radio Telescopes},   {\bf 75}, 82--89  (1995).

\bibitem{candotti10design}
M.~Candotti, ``Design of the alma band 10 optics with horn-to-horn local oscillator signal power injection scheme,'' tech. rep., ALMA.
\newblock [doi:10.13140/RG.2.1.3579.9121].

\bibitem{ALMAhb2019}
ALMA, {\em ALMA Cycle 7 Technical Handbook}, https://almascience.nrao.edu/news/alma-cycle-7-science-observations-status-update-3  (2019).

\bibitem{schwan2011invited}
D.~Schwan, P.~A. Ade, K.~Basu, {\em et~al.}, ``Invited article: Millimeter-wave bolometer array receiver for the atacama pathfinder experiment sunyaev-zel'dovich (apex-sz) instrument,'' {\em Review of Scientific Instruments} {\bf 82}(9)  (2011).
\newblock [doi:10.1063/1.3637460].

\bibitem{Kraus1986}
J.~D. Kraus, {\em Radio Astronomy}, Cygnus-Quasar Books, 2nd~ed.  (1986).

\bibitem{ASC}
{ALMA web page}, ``\texttt{https://almascience.eso.org/proposing/ \\ sensitivity-calculator},''  (12 de {J}une de 2024).

\bibitem{holdaway2001alma}
M.~Holdaway, ``Alma memo 402 illumination taper misalignment and its calibration,''  (2001).

\bibitem{36FootTelescope}
B.~L. Ulich, ``Pointing characteristics of the 36-foot telescope,'' tech. rep., National Radio Astronomy Observatory Green Bank 1 West Virginia  (1976).

\bibitem{cwik1996beam}
T.~Cwik and V.~Jamnejad, ``Beam squint due to circular polarization in a beam-waveguide antenna,'' {\em Telecommunications and Data Acquisition Progress Report} {\bf 128}, 1--10  (1996).

\bibitem{AlmaFeReq}
C.~T. Cunningham, G.~H. Tan, H.~Rudolf, {\em et~al.}, ``Front-end sub-system for the 12 m-antenna array technical specifications alma-40.00.00.00-001-a-spe,''  (2007).

\bibitem{prescott1965monte}
P.~Prescott, ``Monte carlo methods,''  (1965).

\bibitem{siegman1986lasers}
A.~E. Siegman, {\em Lasers}, University science books  (1986).

\bibitem{lazareff2001alignment}
B.~Lazareff and S.~Sakamoto, ``Alignment tolerances for alma optics,'' {\em ALMA Memo 395}   (2001).

\bibitem{rigby2019distributions}
R.~A. Rigby, M.~D. Stasinopoulos, G.~Z. Heller, {\em et~al.}, {\em Distributions for modeling location, scale, and shape: Using GAMLSS in R}, CRC press  (2019).
\newblock [doi:10.1201/9780429298547].

\bibitem{withington1992design}
S.~Withington, J.~A. Murphy, A.~Egan, {\em et~al.}, ``On the design of broadband quasioptical systems for submillimeter-wave radio-astronomy receivers,'' {\em International journal of infrared and millimeter waves} {\bf 13}, 1515--1537  (1992).
\newblock [doi:10.1007/BF01009233].

\end{thebibliography}
\bibliographystyle{spiejour}   

Biographies of authors

\vspace{2ex}\noindent\textbf{Emiliano Rasztocky} Graduated in Mechanical Engineering in 2002 from the Universidad Nacional de La Plata (UNLP). AIV and design engineer for the development of the tertiary optical system for the LLAMA project. Since 2021, system engineer for the QUBIC project. From 2014 to 2024, Chief of the Mechanical Department at the Instituto Argentino de Radioastronomía (IAR). Since 2024, design engineer for the opto-mechanical subsystems for PMI project at the Max Planck Institute for Solar System Research (MPS).

\vspace{2ex}\noindent\textbf{Matías Rolf Hampel} Received his Ph.D. in Engineering from Universidad Tecnológica Nacional (UTN), Argentina, in 2018. He was an international excellence fellow at the Karlsruhe Institute of Technology (KIT), Germany. He is a researcher at CONICET and CNEA, Argentina, and leads the Detector Design and Construction Department at GAIDI-CNEA. He is a member of the Adviser Committee of ITeDA, the Pierre Auger Collaboration, and the QUBIC Collaboration.

\vspace{2ex}\noindent\textbf{Rodrigo Reeves} Received a Ph.D. in electrical engineering from Universidad de Concepcion (UdeC), Chile, in 2009. Served as Chief Engineer at the Chajnantor Observatory from 2005 to 2009, overseeing CBI, CBI2, and QUIET projects. Postdoc in astronomical instrumentation at Caltech, later a Senior Scientist there from 2009 to 2014. Currently an Associate Professor at UdeC, also directing CePIA, the Center for Astronomical Instrumentation. Specializes in cryogenic coherent receivers, microwave array receivers, and sub-millimeter instrumentation with radio-astronomy applications.

\vspace{2ex}\noindent\textbf{Jacques R.D. Lepine} Graduated in Physics (1967), Master in Nuclear Physics (1970)  at Physics Institute of University of São Paulo. PhD in Astronomy at University of Paris VII (1977). Full Professor at Institute of Astronomy, Geophysics and Atmospheric Sciences (IAG-USP). Was president of the Brazilian Astronomical Society(1988-1989), Chair of Astronomy Department (1994-1988), Director of IAG-USP(2001-2005) ,  member of the CNPQ Adviser Committee for Physics and Astronomy. Member of  Brazilian Academy of Sciences, and  World Academy of Sciences.

\vspace{2ex}\noindent\textbf{Gustavo Esteban Romero} 
Received his Ph.D. in Physics from the Universidad Nacional de La Plata (UNLP). He was a postdoctoral fellow at the University of Sao Paulo and a research associate at the Max Planck Institute for Nuclear Physics (Heidelberg, Germany). Currently, he is a Senior Researcher at CONICET, Director of the Instituto Argentino de Radioastronomía (IAR), and a Full Professor at UNLP. He has lectured globally and supervised around 40 doctoral and undergraduate theses.



\end{document}